\documentclass[fleqn,usenatbib]{mnras}

\usepackage[normalem]{ulem}

\usepackage[T1]{fontenc}
\usepackage{gensymb}

\usepackage{graphicx}	% Including figure files
\title[The galaxy environment in GAMA G3C groups using the Kilo Degree Survey Data Release 3]{The galaxy environment in GAMA G3C groups using the Kilo Degree Survey Data Release 3}

\author[]{
M. V. Costa-Duarte $^{1,2}$\thanks{E-mail: mvcduarte@usp.br}, 
M. Viola$^{2}$, 
A. Molino$^{1}$, 
K. Kuijken$^{2}$,
L. Sodr\'e Jr.$^{1}$, \newauthor  
M. Bilicki$^{2,3}$, 
M. M. Brouwer$^{2}$
H. Buddelmeijer$^{2}$,
F. Getman$^{4}$,
A. Grado$^{4}$, \newauthor
J. T. A. de Jong$^{2}$,
G. V. Kleijn$^{6}$,
N. Napolitano$^{4}$,
E. Puddu$^{4}$, 
M. Radovich$^{5}$, 
M. Vakili$^{2}$ \\
$^{1}$Instituto de Astronomia, Geof\'isica e Ciencias Atmosf\'ericas, University of S\~ao Paulo, R. do Mat\~ao 1226, 05508-090 S\~ao Paulo, Brazil\\
$^{2}$Leiden Observatory, Leiden University, P.O. Box 9513, NL-2300 RA, Leiden, The Netherlands \\
$^{3}$National Centre for Nuclear Research, Astrophysics Division, P.O. Box 447, PL-90-950 Lodz, Poland \\
$^{4}$INAF - Osservatorio Astronomico di Capodimonte, via Moiariello 16, 80131 Napoli, Italy \\
$^{5}$INAF - Osservatorio Astronomico di Padova, via dell'Osservatorio 5, 35122 Padova, Italy \\
$^{6}$ Kapteyn Astronomical Institute, University of Groningen, Postbus 800, 9700 AV, Groningen, The Netherlands
}

\date{Accepted XXX. Received YYY; in original form ZZZ}

\pubyear{2017}

\begin{document}
\label{firstpage}
\pagerange{\pageref{firstpage}--\pageref{lastpage}}
\maketitle

% Abstract of the paper
\begin{abstract}

We aim to investigate the galaxy environment in GAMA Galaxy Groups Catalogue (G3C) using a volume-limited galaxy sample from the Kilo Degree Survey Data Release 3. The k-Nearest Neighbour technique is adapted to take into account the probability density functions (PDFs) of photometric redshifts in our calculations. This algorithm was tested on simulated KiDS tiles, showing its capability of recovering the relation between galaxy colour, luminosity and local environment. The characterization of the galaxy environment in G3C groups shows systematically steeper density contrasts for more massive groups. The red galaxy fraction gradients in these groups is evident for most of group mass bins. The density contrast of red galaxies is systematically higher at group centers when compared to blue galaxy ones. In addition, distinct group center definitions are used to show that our results are insensitive to center definitions. These results confirm the galaxy evolution scenario which environmental mechanisms are responsible for a slow quenching process as galaxies fall into groups and clusters, resulting in a smooth observed colour gradients in galaxy systems. 

\end{abstract}


\section{Introduction}

The hierarchical structure formation theory predicts that the primordial density field in the early Universe evolves through gravitational instabilities and its final stage is represented by virialized dark matter dominated haloes. These systems also represent potential wells for the baryonic matter, which is gravitationally trapped, allowing galaxies to form \citep{WhiteRees1978}. Additionally, galaxies tend to cluster into larger structures and form the so-called cosmic web \citep{Vogeleyetal2004, GottIIIetal2005}. Several works have shown that the environment within galaxy systems is essentially responsible for the galaxy quenching. Red galaxies are more often found in the densest regions of triplets, groups, clusters \citep{Tempeletal2012, CostaDuarteetal2016} and superclusters of galaxies \citep{Einastoetal2011,CostaDuarteetal2013, Einastoetal2014}. From an observational point of view, some galaxy properties are strongly correlated to their local environment, such as colours, stellar population ages and morphology. It can be observed \textit{in situ} in the local Universe and its consequences are the morphology-density and colour-density relations \citep{Dressler1980, Gotoetal2003, Kauffmannetal2004, Dressleretal2013}. In the colour-magnitude diagram (CMD), the mean colour of galaxies is independent of environment, but the red galaxy fraction increases as the local density increases at fixed luminosity \citep{Baloghetal2004a, Balletal2008}. The galaxy colours seem to be more correlated to the environment than the morphology \citep{Kauffmannetal2004, Quinteroetal2006, MartinezMuriel2006}, indicating that the morphological transformation is a subsequent and slower process.  

The galaxy-galaxy and galaxy-cluster interactions are the main responsible mechanisms for the observed star formation quenching over a wide redshift range. Several mechanisms are candidates for the galaxy quenching in distinct regions of galaxy clusters, such as merging \citep{Icke1985, Mihos1995}, harassment episodes \citep{Mooreetal1996, Mooreetal1999}, strangulation \citep{Larsonetal1980, Bekkietal2002} and ram-pressure \citep{GunnGott1972}. The role and contribution of each environmental mechanisms for the quenching process and stellar mass build-up still remains unclear \citep[e.g.,][]{Capaketal2007, vanderWeletal2010, Rowlandsetal2018}. This current scenario suggests a slow gas removal from late-type galaxy haloes with no observed structural changes, with galaxies becoming quiescent due to the lack of gas reservoir for star formation but keeping their morphology still disk-like. Afterwards, a morphological transformation takes place due to more significant gravitational interactions at inner cluster regions and finally elliptical and red galaxies (so-called \textit{red and dead}) mostly populate central regions of galaxy clusters \citep{Baloghetal1998}.

Beyond the local Universe (z$\sim$1), \cite{Sobraletal2011} showed that the stellar mass is the main parameter driving the galaxy quenching, however, the environment also enhances the star formation rate of low-mass objects but quenches all galaxies located at high density regions (groups and clusters). On the other hand, some authors found a positive correlation between the star formation rate and the environment \citep{Elbazetal2007, Cooperetal2008, Tranetal2010, Allenetal2016}, the opposite as found at low redshifts \citep[e.g.][]{Baloghetal2004a}. At z$\sim$1-2, field galaxies present redder colours and lower star formation rates when compared with cluster members \citep{Grutzbauchetal2011, Quadrietal2012}. These results suggest that the environment already plays a significant role between 1$<$$z$$<$2 although when, where and how it specifically starts affecting galaxies is still unclear.     

Several techniques have been proposed to measure and quantify galaxy environment. Among an assortment of methods in the literature, the $k$-Nearest Neighbours ($k$-NN) is widely employed for statistical learning in Astronomy \citep{Hestieetal2009, Kugleretal2015}, including density field reconstruction. This technique calculates the local density of galaxies using the sky projected or 3D distance from a certain galaxy which englobes the $k$ nearest neighbours \citep{Mateusetal2004, Baldryetal2006, Haasetal2012}. In addition, the robustness of recovering galaxy environment is tightly correlated to the redshift precision, which provides the distance between galaxies in the line-of-sight direction. Modern galaxy surveys can usually only provide one of two different redshift measures, spectroscopic and photometric ones. Spectroscopic redshifts (hereafter spec-zs) usually have high precision, but are observationally time consuming and its galaxy sample often suffers from incompleteness. Photometric redshifts (hereafter photo-zs) can be considered as an alternative choice to overcome these issues, however, their precision is systematically lower, mainly depending on the wavelength range and photometric signal-to-noise ratio. Adapted techniques have been proposed to estimate the galaxy environment and the density field, taking into account the photo-z uncertainties \citep{Scovilleetal2013, Malavasietal2016, Malavasietal2017}. Other works have also evaluated the influence of the photo-z uncertainties for future photometric surveys \citep{EtheringtonThomasetal2015, Cucciatietal2016, Laietal2016}. Moreover, the two-point correlation function has been successfully obtained from photo-z galaxy samples \citep[e.g.][]{Soltanetal2015, Asoreyetal2016}. Nowadays, numerous techniques are adapted to include photo-zs in their calculations and to evaluate the role of environment in the galaxy evolution context.

The local density of galaxies is equated with galaxy environment here. As a certain galaxy system (e.g. galaxy clusters, filaments or voids) can present a wide range of values for local density, we understand that the local density traces better the local environment than the structure itself.   

In this paper, we investigate the galaxy environment in the G3C groups. We take the advantage of the KiDS imaging survey to carry out a homegeneous analysis of the galaxy population in these systems. This paper is organized as follows. In Section \ref{sec.data}, we present the databases and their extracted samples used in this work. Section \ref{sec.gal_environment_technique} describes the adapted $k$-NN technique employed to estimate the galaxy environment and the necessary adjustments for the photo-zs of our galaxy samples. In Section \ref{sec.galenv_results}, we show the performance of the adapted $k$-NN technique and the main results obtained from the KiDS/DR3 galaxy sample. Section \ref{sec.g3c_groups_results} presents our main results on the galaxy population in GAMA/G3C groups. Finally, Appendix \ref{app.sec_center_influence} shows the influence of the group center definition on our results. Throughout this paper, we assume the $\Lambda$CDM cosmology with parameters ($h$, $\Omega_m$, $\Omega_{\Lambda}$, $\Omega_{k}$) = (0.7, 0.3, 0.7, 0.0). 

\section{DATA}
\label{sec.data}

\subsection{KiDS Data Release 3}
\label{subsec.kidsdata}

The Kilo Degree Survey (KiDS) performs deep imaging using four photometric broad bands ($ugri$) and covers sky regions in both hemispheres using the VLT Survey Telescope\footnote{http://www.eso.org/public/teles-instr/paranal-observatory/vlt/}. There are two main observed patches, being KiDS North close to the celestial Equator and KiDS South around $\delta$$=$$-31^{\degree}$. The photometric depths for $u$, $g$, $r$ and $i$ are 24.3, 25.1, 24.9, 23.8 (AB magnitudes within 5$\sigma$), respectively. Recently, the KiDS survey presented the Third Data Release to the community \citep[DR3,][]{deJongetal2017}. The KiDS/DR3 is composed of 440 tiles, each one covering 1 sq.deg.

\begin{figure}
\centering
\includegraphics[scale=0.55]{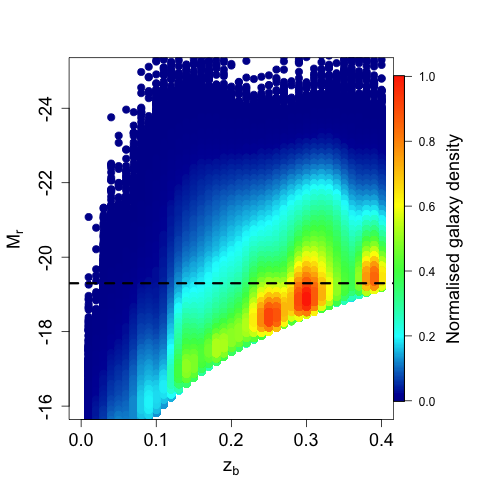}
\caption{The volume-limited sample extracted from the KiDS/DR3 database. The dashed line represents the luminosity threshold imposed to the sample. The high-density regions at z$\sim$0.25, 0.3 and 0.4 can probably be redshift artifacts or overdensity regions. Note that we do not consider most of galaxies from these regions in our analysis.}
\label{fig.vol_limited}
\end{figure}

The KiDS multi-band catalogs provide flags that classify objects as stars or galaxies, and indicate their photometric quality, both generated by the software \texttt{Pulecenella} \citep[][for more details]{deJongetal2015}. The flag \texttt{2DPHOT} represents a morphology-based star/galaxy classification and \texttt{IMAFLAG\_ISO} indicates if the photometry is contaminated by observational issues (bad pixels, cosmic rays, saturated stars, etc). The galaxy sample used in this work is extracted from the multi-band catalog, with \texttt{2DPHOT}=0 and \texttt{IMAFLAG\_ISO\_r}=0. These flags mean that objects are reliably classified as galaxies and have accurate photometry. The \texttt{MAG\_AUTO} magnitudes are corrected by the Galactic extinction and homogenized using the zero-point offsets provided for each photometric band and tile\footnote{http://kids.strw.leidenuniv.nl/DR3/format.php}. Because of the photometric depth at $r$ band slightly varies among KiDS tiles, a conservative magnitude limit is adopted by selecting objects brighter than $r$=22.5. The package \texttt{kcorrect} v4.2 \citep{Blantonetal2007} is employed to obtain the rest-frame magnitudes using the photo-z provided by the \texttt{BPZ} code \citep{Benitez2000}. We extract volume-limited galaxy samples from each KiDS/DR3 tile, taking objects brighter than $M_{\rm r}$$<$-19.3 and within the redshift range 0.01$<$$z$$<$0.4. Figure \ref{fig.vol_limited} shows the extracted samples from the KiDS/DR3 database. In total, our final sample consists of 1080224 galaxies distributed into 440 tiles. In our analysis, we consider the following luminosity bins: -20$\le$$M_{\rm r}$$<$-19.3, -21$\le$$M_{\rm r}$$<$-20, -22$\le$$M_{\rm r}$$<$-21 and -23.5<$M_{\rm r}$$<$-22. The brightest luminosity bin is limited up to -23.5 in order to avoid outliers in our sample. 

Some observational effects, such as bad pixels, saturated stars and cosmic rays are often responsible for unreliable photometry of sources and also compromise the sky continuity in our analysis. The mean fraction of these regions over all KiDS/DR3 tiles is 17\%. In order to track these problematic regions, objects presenting \texttt{IMAFLAG\_ISO}$>$0 can be used as tracers for these issues. The affected sky area can be estimated using the number of sources with \texttt{IMAFLAG\_ISO}$>$0 over the total number of objects within a certain region. Sources up to $r$=25 and with flag \texttt{IMAFLAG\_ISO} are extracted as auxiliary tile samples. Note that these auxiliary samples are only included in our calculations to map observational issues which affect the projected galaxy distribution. 

%\begin{figure*}
%\centering
%\includegraphics[scale=0.55]{tiles_KiDS_DR3_north_list.png}
%\includegraphics[scale=0.55]{tiles_KiDS_DR3_south_list.png}
%\caption{The KiDS/DR3 tiles and their disposition in the sky. Upper panel: KiDS-North tiles. Lower panel: KiDS-South tiles.}
%\label{fig.tiles_KiDS}
%\end{figure*}

\subsection{KiDS Mock Catalog}
\label{subsec.mock_cat}

\begin{figure}
\centering
\includegraphics[scale=0.45]{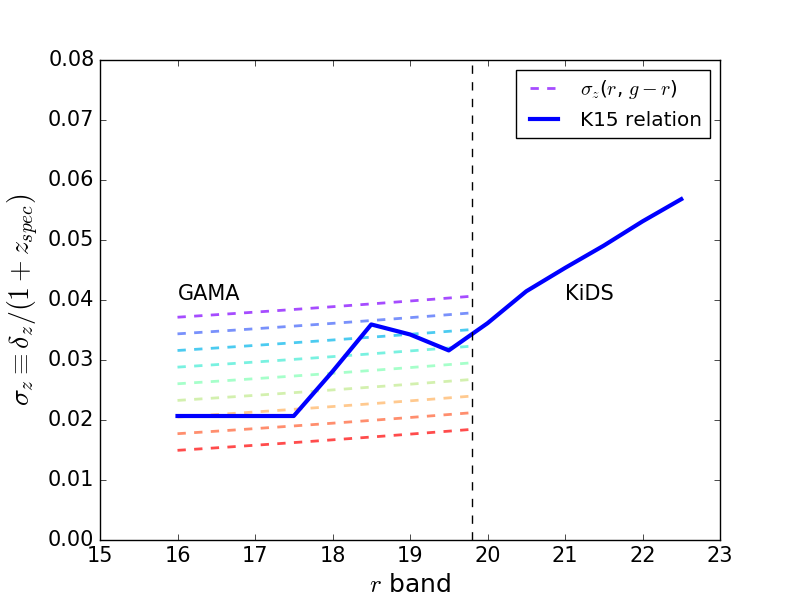}
\caption{The redshift uncertainties ($\sigma_z$) as a function of $r$-band and $(g-r)$. The blue continuous line is $\sigma_z$ as a function of $r$ band, following the photo-z uncertainty behaviour from Kuijken et al.(2015). Dashed lines represent the $(g-r)$-dependent redshift uncertainties obtained from the match between GAMA and KiDS surveys. Lower dashed parallel lines represent redder galaxies.}
\label{fig.sigmaz_KiDS_GAMA}
\end{figure}

To evaluate the capability of the proposed galaxy environment algorithm (see section \ref{sec.gal_environment_technique}), we generate KiDS mock catalogs. The photo-z modelling on the KiDS mock catalog consists of two parts. The behaviour of the photo-zs as a function of: i) the apparent magnitude ($r$ band), and ii) the galaxy colour $(g-r)$. The first one is obtained from Fig. 11 of \cite{Kuijkenetal2015}(hereafter K15), using the relation between $\sigma_{\rm z}\equiv std(\delta_{\rm z}/(1+z_{\rm spec}))$ and the apparent $r$ band, where $\delta_{\rm z} = (z_{\rm phot} - z_{\rm spec})$. The second one requires spectroscopic data to evaluate the photo-z uncertainties as a function of galaxy colours. The spectroscopic GAMA survey \citep{Driveretal2009} overlaps with the KiDS coverage for four fields (G09, G15, G12 and G23), however, presenting a shallower sample ($r$$<$19.8) than the KiDS photometric depth. The match between KiDS and GAMA samples within 1 arcsec provided roughly 168k objects. The photo-z uncertainty is then modeled as a function of the galaxy colour and apparent magnitude ($\sigma_{\rm z}(r, g-r)$) up to $r$=19.8. For fainter objects, the relation presented by K15 is adopted. Figure \ref{fig.sigmaz_KiDS_GAMA} shows the photo-z uncertainties as a function of the $r$-band and galaxy colour $(g-r)$ for the bright side and the relation from K15 for fainter objects. The representative photo-z uncertainty is obtained as the mean value over the apparent magnitude range, being 0.042(1+$z$).

Our KiDS/DR3 sample also suffers from photo-z outliers. Particularly, high redshift galaxies (0.4$<$$z_{\rm spec}$$<$1.4) can have their photo-zs wrongly assigned between 0.0$<$$z_{\rm b}$$<$0.4 by the BPZ code. \cite{deJongetal2017} shows a matched galaxy sample between the KiDS/DR3 and the zCOSMOS survey \citep{Lillyetal2007} and its outliers ($|z_{\rm phot} - z_{\rm spec}|$/(1+z$_{\rm spec}$)$>$0.15) (see their Figure 11). According to \cite{Bilickietal2017}, this fraction represents 8\% of galaxies up to $r$=25. The photo-z outliers are modeled in mock catalogs using their observed outlier photo-z distribution between 0.01$<$$z_{\rm b}$$<$0.4 \citep{deJongetal2017}. In our mock catalog, galaxies with $z_{\rm spec}>0.4$ are namely brought to the redshift range of our sample following the outlier fraction and photo-z distribution. In this way, mock outliers reproduce the observed fraction and distribution of KiDS/DR3 database. 

We extract eight simulated KiDS-like tiles from the mock lightcones produced by \cite{Mersonetal2013}\footnote{http://www.star.ucl.ac.uk/~aim/lightcones.html}. Modeling of the photo-z and its outliers is also performed. The probability density functions (PDF(z)) are reconstructed following the central value ($z_{\rm b}$) and 95\% of confidence level limits ($z_{\rm b,min}$ and $z_{\rm b,max}$), i.e., using Gaussian gradient approximation which follow the confidence levels $z_{\rm b,min}$ and $z_{\rm b,max}$. We define volume-limited samples from these simulated tiles, extracting objects brighter than $M_{\rm r}$$<$-19.3 in the redshift range 0.01$<$$z_{\rm b}$$<$0.4. The final simulated sample had 18964 galaxies distributed over eight KiDS tiles. Areas affected by observational issues are modeled following the observed 17\% of problematic regions, as described in subsection \ref{subsec.kidsdata}. Regions randomly distributed over all simulated tiles had their sources flagged as \texttt{IMAFLAG\_ISO}$>$0 up to $r$$<$25 to mimic the pattern found in the observed KiDS tiles. 

\begin{figure*}
\centering
\includegraphics[scale=0.9]{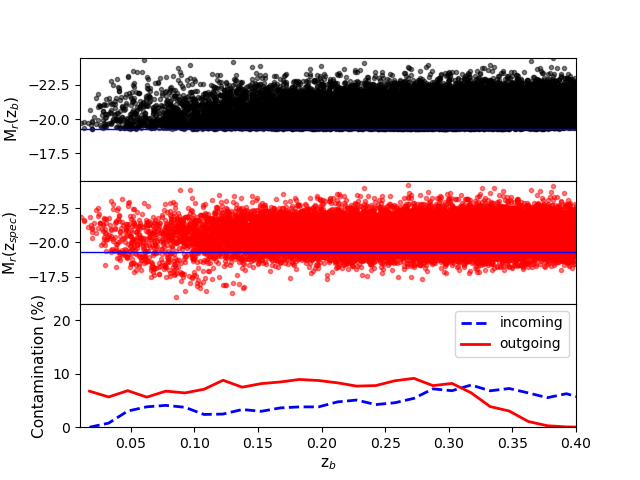}
\caption{Simulated volume-limited sample and the contamination due to photo-z uncertainties. Upper: sample initially constrained using the absolute magnitudes calculated from photo-zs. Middle: galaxy sample from the upper panel but showing absolute magnitudes calculated using spec-zs. Lower: Contamination fraction of galaxies as function of the redshift. Solid and dashed lines represent the incoming and outgoing over all simulated KiDS tiles, respectively.}
\label{fig.contamination_zb}
\end{figure*}

The photo-z uncertainties surely affect the absolute magnitudes and galaxies can be wrongly included or excluded from our volume-limited sample, according to their photo-z uncertainties. Overestimated photo-zs can include galaxies which are fainter than -19.3 into our volume-limited sample, being an incoming. On the other hand, underestimated photo-zs can exclude objects brighter than the absolute magnitude limit from the sample, i.e., outgoing. Figure \ref{fig.contamination_zb} shows the mock volume-limited sample extracted using photo-zs (upper panel) and the same sample by considering absolute magnitudes from the spec-zs (middle panel). In the lower panel, we show the incoming and outgoing as a function of the photo-z. The incoming presents constant values (around 7\%) and decreases as it gets closer to the redshift limit (z$_{\rm b}$=0.4). The incoming reaches null value at the upper redshift limit due to the magnitude cut $r$=22.5. It excludes all objects fainter then -19.3 at the redshift limit then the incoming is null at this redshift by definition. The incoming is null at the lower redshift range since it comes from underestimated photo-zs. The outgoing roughly increases with photo-z due to the photo-z uncertainty being larger at higher photo-zs. The contamination parameters are further employed to correct the local density (see Section \ref{sec.gal_environment_technique}).

\subsection{GAMA/G3C catalog}
\label{subsec.gama_g3c}

The GAMA project \citep[Data Release 2,][]{Liskeetal2015} is an extragalactic multiwavelength survey which combines photometry (far-UV to radio) and optical spectra of more than 290000 objects over $~$286 sq.deg. The optical spectroscopy employs the AAOmega spectrograph on the Anglo-Australian Telescope (AAT). The aperture matched photometry provides optical SDSS petrosian magnitudes ($ugriz$) and infrared bands ($YJHK$) from the VIKING survey \citep{Edgeetal2013} for targets down to $r_{AB}$=19.8. An impressive spectroscopic completeness ($\sim$98\%) provides a database which allows to investigate several topics in galaxy evolution, such as the galaxy environment, stellar populations and halo formation times.  

The G3Cv5 \citep{Robothametal2011} is a galaxy group catalog of which the GAMA/G15 patch is publicly available. It is worth to mention that G3C groups means all galaxy systems with more than 5 members. The \textit{friends-of-friends} (FoF) algorithm was adapted to take into account the selection function of the survey to identify galaxy systems up to $z$$<$0.5. Their halo masses $\log_{10}M_{\rm h}$ (parameter $MassA$) were estimated from a dynamical proxy, using the group velocity dispersion ($\sigma_{\rm group}$) and the projected radius which contains 50\% of the members ($R_{\rm 50}$) as well as the scaling factor A \citep[for more details, see][]{Robothametal2011}. Galaxy groups more massive than $\log_{10}M_{\rm h}$$=$13.21 that have five or more members and are located between 0.01$<$$z$$<$0.4 are considered in our analysis, resulting in 348 galaxy systems in total. We divide our G3C sample into three sub-samples according to their masses: 13.21$<\log_{10}(M_{\rm h})\le$13.69 (G1), 13.69$<\log_{10}(M_{\rm h})\le$14.05 (G2), $\log_{10}(M_{\rm h})>$14.05 (G3), corresponding to 116 galaxy systems for each bin. It is worth to note that, since our analysis does not compare group properties at different redshifts, it is not necessary to have a group volume-limited sample. Assuming a mass threshold for galaxy groups and the galaxy volume-limited sample, it ensures that the galaxy luminosity threshold is the same over the entire redshift range and a homogeneous galaxy population analysis can be carried out over the aforementioned redshift range. We initially adopt the brightest group galaxy as group center (denominated BCG in the GAMA G3C catalog). The influence of center definition on our results is also evaluated, using an alternative center definition: the r-band luminosity weighted center.

\section{The galaxy environment}
\label{sec.gal_environment_technique}

\subsection{k-Nearest Neighbour technique (k-NN)}

The majority of the works on galaxy environment relying on photo-zs does not consider PDF(z) in their calculations, or even include a simplified version of PDFs using $z_{\rm b,min}$ and $z_{\rm b,max}$ (95\% of confidence level). The influence of the PDFs in the galaxy environment estimates is evaluated in our simulations and the results are compared to the original k-NN technique in spec-z space. 

The initial algorithm considers a certain galaxy in the sky at $R_0$ and redshift $z_0$. The algorithm defines a projected radius $R_{\rm kNN}$ around this galaxy which encloses the $k$ nearest projected galaxies in the sky within the redshift range $z_{\rm 0}\pm\Delta_{\rm z}(1+z_{\rm 0})$, where $\Delta_{\rm z}$ is the average redshift uncertainty of the survey. The surface density is then defined as follows,
\begin{equation}
\label{eq.knn_original}
\centering
\sigma_{\rm kNN}(R_0, z_0) = \frac{k}{\pi R_{\rm kNN}^2}.
\end{equation}
Notice that the length of the cell defined above follows the redshift uncertainty of the survey, i.e., it includes the term $(1+z)$ in the calculations. 

The adaptation of the aforementioned algorithm for PDF inclusion substitutes the number $k$ by summing the galaxy probabilities of being within the redshift interval $z_0\pm\Delta_z(1+z_0)$ until reaching the number of neighbours $k$. As the PDFs provide probabilities in redshift space, the concept of neighbours is now adapted in a probabilistic formalism. The probability of the i-th galaxy of being in the redshift range is, 
\begin{equation}
\label{eq.integral_PDF}
\centering
P_i = \int^{z_0+\Delta_z(1+z_0)}_{z_0-\Delta_z(1+z_0)} PDF(z) dz.
\end{equation}
The local density of galaxies with the inclusion of PDFs can be written as, 
\begin{equation}
\label{eq.knn_PDF}
\centering
\sigma(R_0, z_0) = \frac{S_k}{\pi R_{\rm kNN}^2}.
\end{equation}
where $S_k$ is the sum of $P_i$ over all galaxies enclosed by $R_{\rm kNN}$. The projected radius $R_{\rm kNN}$ increases until the probability sum of all neighbour candidates reaches the desired value $k$ ($S_k=k$) in contrast to solely galaxy counting, as shown in the initial technique. 

The evaluation of both algorithms described above in the mock catalogs is carried out using the k-NN technique in spec-z space. In this case, the local densities are estimated using spheres with radius $r_{\rm KNN}$ which contain $k$ nearest neighbours. The spectroscopic local density is then defined as $\rho_{\rm spec}(R_0, z_0) = \frac{k}{(\frac{4}{3} \pi r_{\rm kNN}^3)}$. This volumetric density is considered as the reference galaxy environment. Although the galaxy environment densities calculated in spec-z and photo-z are not the same by definition, we are able to find a positive correlation between them. 

\subsection{Contamination, Masking and Border effects}
\label{subsec.maskborder}

The galaxy environment formalism presented above is still affected by contamination, masking and border effects. Some galaxies located at the tile border can have their local density underestimated due to the non-continuity of the survey. The correction needed to this missing area is defined as the fraction of the circle projected in the sky with radius $R_{\rm kNN}$ situated outside of the survey boundaries or affected by bad pixels, i.e., $f_{\rm area} = A_{\rm out} / \pi R_{\rm kNN}^2$. This method assumes that the area outside the circle presents the same local density of galaxies obtained within the survey area. The area correction weight is then defined as $w_{\rm area} = 1 / (1 - f_{\rm area})$ and it increases as the missing area fraction increases. If there is no missing area, $f_{\rm area}=0$ and consequently $w_{\rm area}=1$. A similar correction is also necessary due to redshift limits of the sample. The individual redshift ranges for each galaxy cell previously defined can have part of its volume outside the redshift range of the galaxy sample and their local densities can be again underestimated. The redshift correction simply consists of the volume fraction outside the survey, $f_{\rm z} = V_{\rm out} / V_{\rm cell}$ and similarly $w_{\rm z}=1/(1-f_{\rm z})$. The sample contamination described in the subsection \ref{subsec.mock_cat} is corrected by using a similar formalism: $w_{\rm C}(z) = (1 - f_{\rm C}(z))$, where $f_{\rm C}(z)$ is the difference between the incoming and outgoing as a function of the photo-z, i.e., $f_{\rm C}(z)=C_{\rm incoming}(z) - C_{\rm outgoing}(z)$. If $C_{\rm incoming}$ is larger/smaller than $C_{\rm outgoing}$, $w_{\rm C}$ is lower/higher than unity. This correction takes into account the galaxy contamination in our volume-limited sample.

The local density of galaxies is simultaneously corrected by sky area, volume and contamination, i.e., $\sigma_{\rm corr}(R_0, z_0) = \sigma(R_0, z_0) w_{\rm area} w_{\rm z} w_{\rm C}$. The volume and area corrections are essentially geometrical and are applied to spectroscopic and photo-z samples following their individual geometry, accordingly. 

As our observed and simulated data is configured in tiles, the tile management is mandatory for the calculations in order to reduce the border effect and consequently increasing the sky continuity. Some of the tiles have neighbouring tiles around them \citep[see Figure 1 from][]{deJongetal2017} while others are basically isolated. Tiles which have others nearby form a larger contiguous area and consequently these close tiles are included in the calculations to maximize the continuity of the sky area. The galaxy environment for isolated tiles are simply calculated without the inclusion of other tiles. It means that the border corrections are frequently applied for these tiles. 

\section{Galaxy environment results}
\label{sec.galenv_results}

\subsection{Simulated KiDS/DR3 Database}
\label{subsec.env.uncertainty}

The local density of galaxies is regularly transformed to density contrast in the literature in order to make comparable different galaxy environment techniques or parametrizations. Hereafter, the local density of galaxies is converted to density contrast as follows,
\begin{equation}
1 + \delta = \frac{\sigma}{\bar{\sigma}},
\end{equation}
where $\sigma$ is the local density and $\bar{\sigma}$ represents the average density. 

Since the galaxy environment technique previously presented is parametrized as a function of the number of neighbours, we adopt values from $k$=2 to $k$=50. The k-NN technique traces galaxy environments at large scales for large values of k, with the environmental scale proportional to the value of k.  

\begin{figure*}
\centering
\includegraphics[scale=0.45]{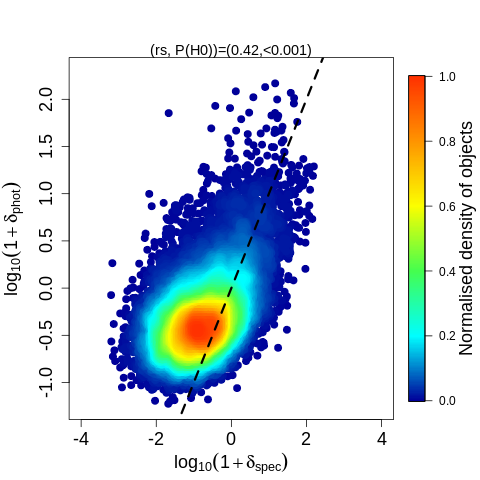}
\includegraphics[scale=0.45]{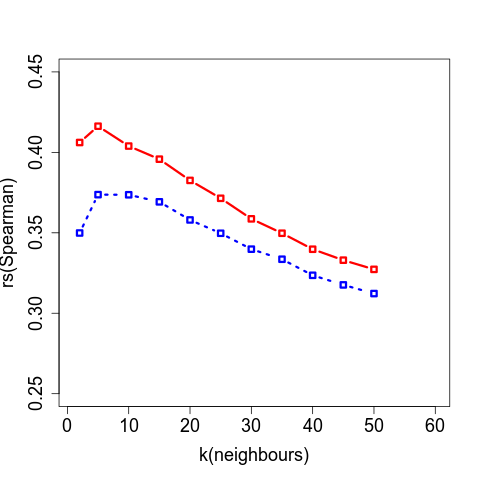}
\caption{Results for eight simulated KiDS tiles. Left: Correlation between the density contrast dereived from spec-z and photo-z using the PDFs for $k$=5. Right: Spearman correlation coefficient as a function of the parameter $k$, i.e., number of the neighbours (see Section \ref{sec.gal_environment_technique}). Solid/dashed lines represent the k-NN technique with/without the inclusion of the reconstructed PDFs in the calculations, respectively.}
\label{fig.rho_spec_phot}
\end{figure*}

The Spearman correlation coefficient evaluates possible correlation between the density contrasts in the spec-z and photo-z spaces from the simulated KiDS/DR3 tiles. This coefficient $r_s$ varies between -1 and +1, indicating anti-correlation and correlation between two sets, respectively. The null hypothesis probability ($P(H_0)$) says how probable these two sets of data are correlated. It also indicates whether $r_s$ is statistically significant or not, preferred to be lower than $10^{-3}$ or $<$3$\sigma$ for a significant correlation. Figure \ref{fig.rho_spec_phot} (left) shows the density contrast comparison between the spectroscopic ($\log_{10}(1+\delta_{\rm spec})$) and photometric ($\log_{10}(1+\delta_{\rm phot})$) redshift spaces. This result indicates that the galaxy environment can be estimated in the KiDS survey using the technique presented in the Section \ref{sec.gal_environment_technique}. Positive correlations are found for both approaches (the inclusion or not of PDFs) and several numbers of neighbours. Note that the relation between the density contrasts is not centred on the 1:1 line. The density field of galaxies follows roughly a log-normal distribution, and any other normalisation which does not use the median value would not bring both distributions centred at (0,0). As our analysis is presented comparatively, it should not affect our results. Figure \ref{fig.rho_spec_phot} (right) also shows the Spearman correlation coefficient as a function of the number of neighbours with and without the PDF inclusion in our calculations. The correlation coefficient peaks at ($r_s$, P(H$_0$))=(0.42,$<10^{-3}$) and $k$=5, and decreases as the number of neighbours increases. We choose $k$=5 for our further analysis in this paper. It seems that low number of neighbours ($k$$<$5) is more susceptible to redshift uncertainties due to the low countings. At larger scales ($k$$>$5), local densities lead to smaller amplitudes of density contrasts since at those scales the Universe is more homogeneous. The optimized number of neighbours is in agreement with previous works for spec-zs surveys \citep{Baloghetal2004a, Baloghetal2004b, Baldryetal2006}. The PDF influence is also evident in this comparison. The Spearman correlation coefficients for results including PDFs are systematically higher than those without PDFs. This difference becomes more evident as the number of neighbours decreases. It shows the importance of the PDFs in our calculations to recover the galaxy density field. 

\subsection{Observed Data}

The galaxy environment technique described in Section \ref{sec.gal_environment_technique} is applied on all tiles in KiDS/DR3 database, covering the number of neighbours from $k$=2 to $k$=50. However, further results are shown only for $k$=5, presenting a relatively higher correlation coefficient. The KiDS/DR3 density contrasts are then divided into quartiles of sources according to their density contrasts: $\log_{10}(1+\delta)$$\le$-0.25, -0.25$<$$\log_{10}(1+\delta)$$\le$-0.11, -0.11$<$$\log_{10}(1+\delta)$$\le$+0.05 and $\log_{10}(1+\delta)$$>$+0.05. These bins are chosen in order to have a significant number of objects in all density contrast bins and roughly separate galaxies into low density, mean density, overdensity and high density environments.

The galaxy classification between red and blue is adopted in further analysis using a $(g-r)$ limit presented by \cite{Cooperetal2010} (C10) to define the blue limit of the red sequence, 
\begin{equation}
(g-r)_{C10}=-0.02667 M_{\rm r}  + 0.11333,
\end{equation}
where $M_{\rm r}$ is the absolute magnitude at $r$ band. Objects above or below this colour and luminosity thresholds are classified as red or blue galaxies, respectively. 

\begin{figure*}
\centering
\includegraphics[scale=0.25]{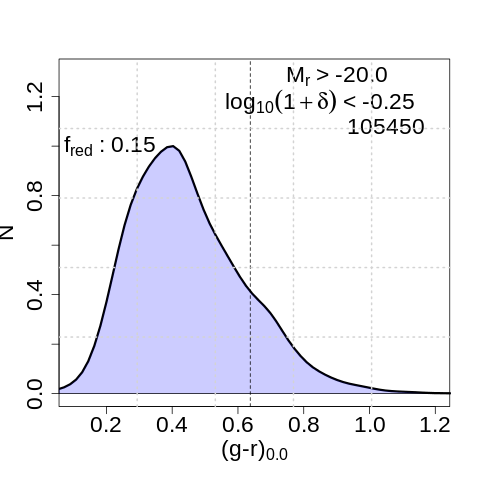}
\includegraphics[scale=0.25]{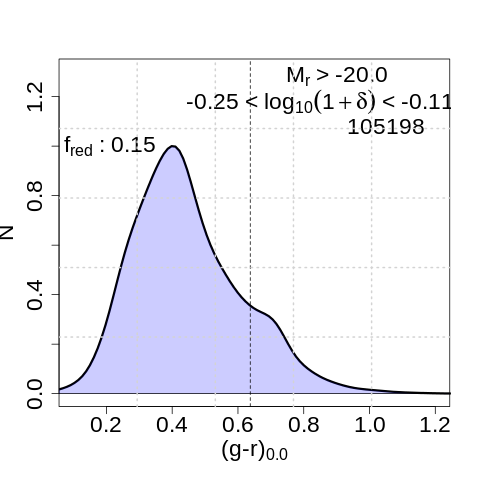}
\includegraphics[scale=0.25]{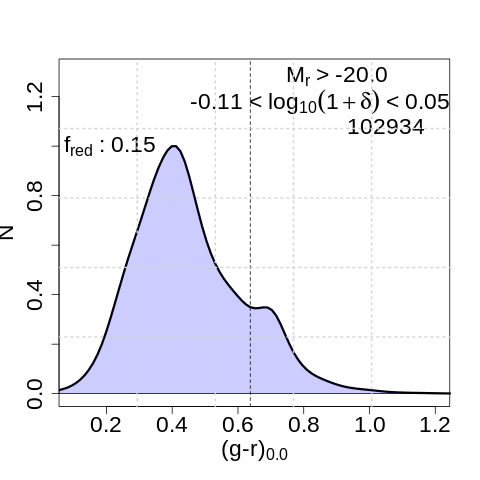}
\includegraphics[scale=0.25]{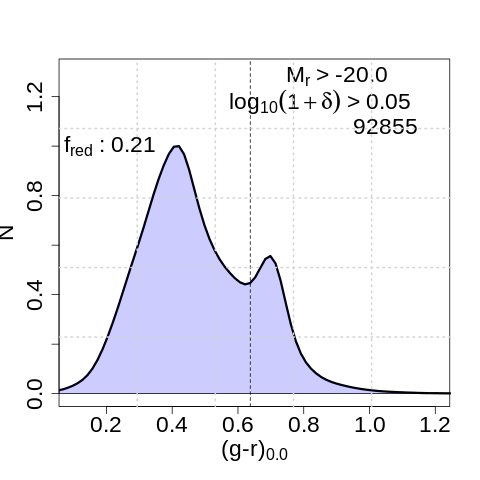}

\includegraphics[scale=0.25]{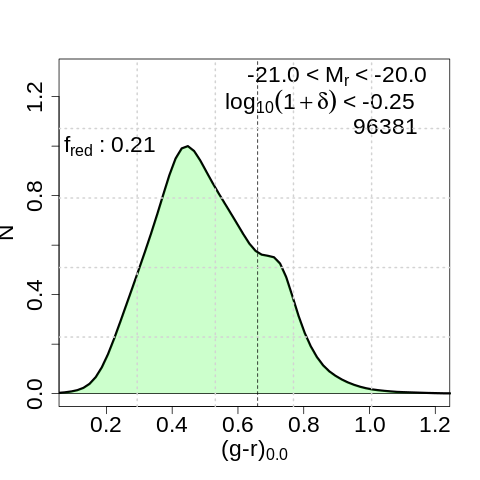}
\includegraphics[scale=0.25]{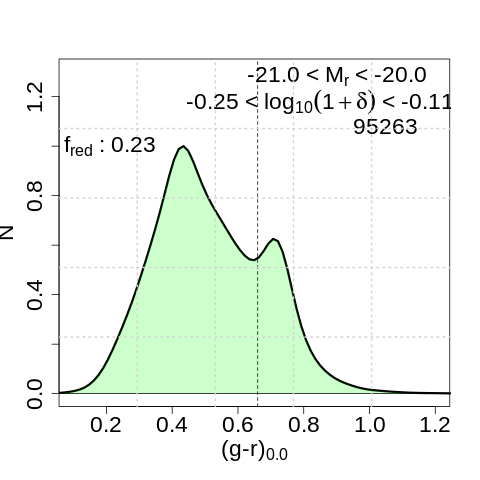}
\includegraphics[scale=0.25]{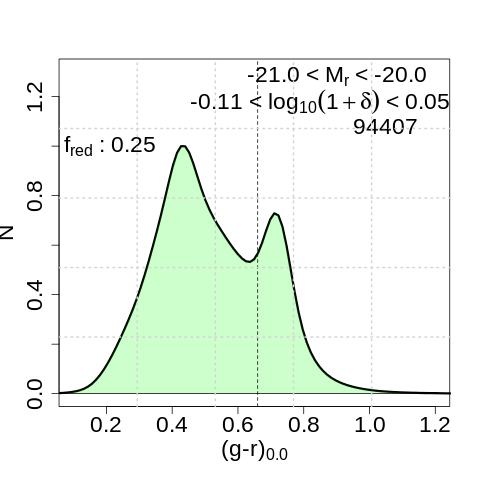}
\includegraphics[scale=0.25]{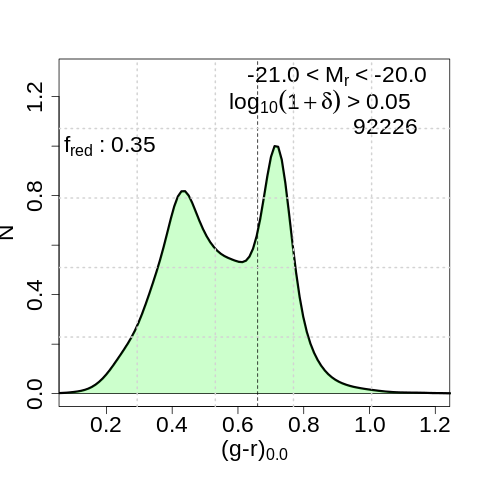}

\includegraphics[scale=0.25]{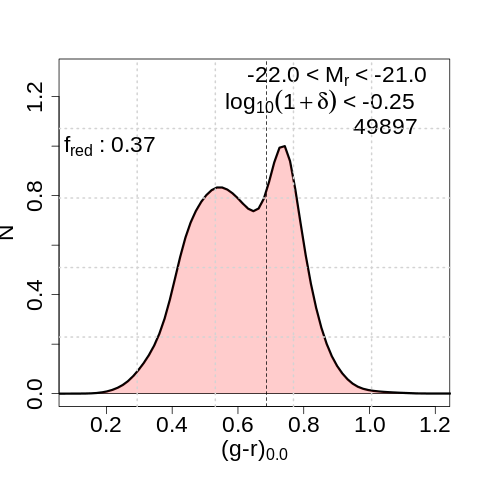}
\includegraphics[scale=0.25]{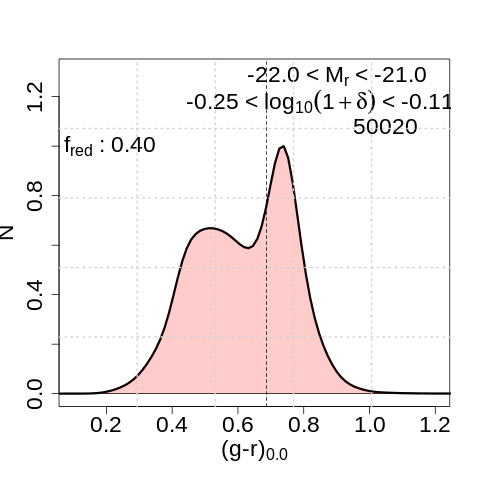}
\includegraphics[scale=0.25]{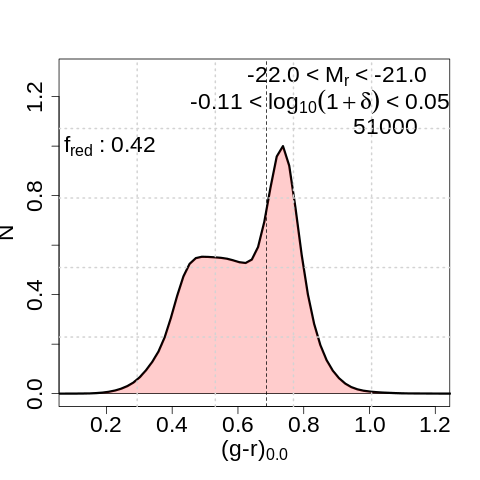}
\includegraphics[scale=0.25]{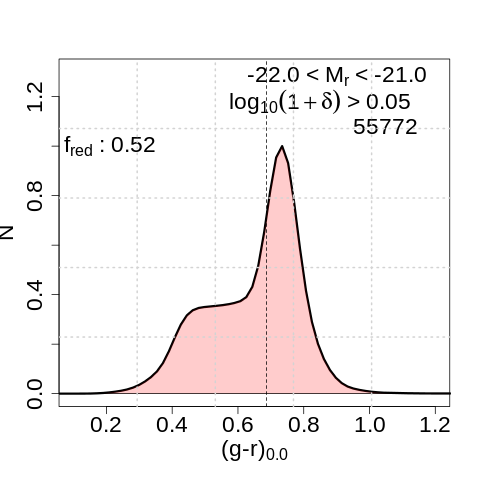}

\includegraphics[scale=0.25]{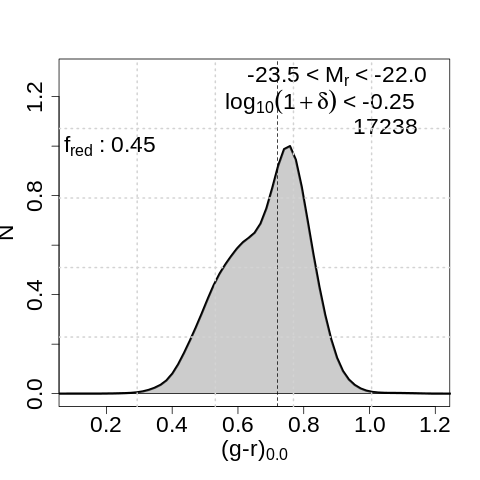}
\includegraphics[scale=0.25]{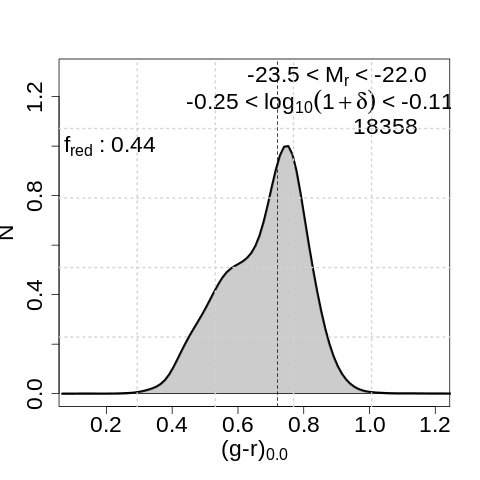}
\includegraphics[scale=0.25]{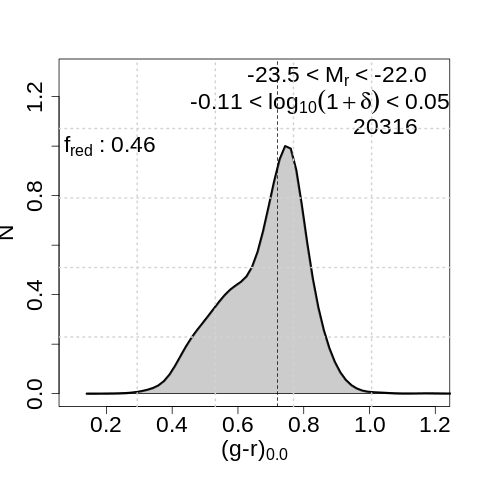}
\includegraphics[scale=0.25]{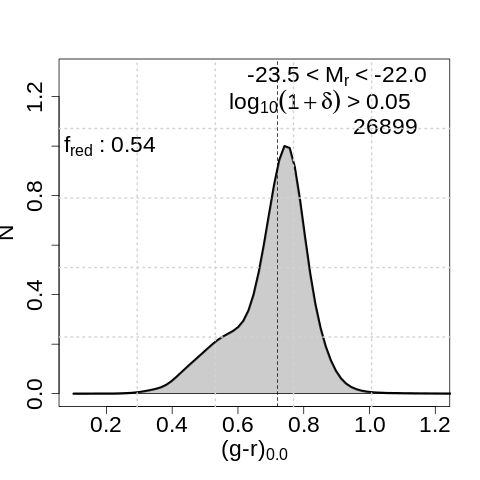}

\caption{Normalised histograms of rest-frame colour $(g-r)_{0.0}$ for the KiDS/DR3 volume-limited sample between 0.01$<$$z_{\rm b}$$<$0.4. The panels show the colour distribution as a function of the local density contrast and absolute magnitude bins. The galaxy luminosity bins (see Section \ref{sec.data}) are represented by blue, green, red and grey distributions, respectively. The dashed vertical lines represent the colour threshold adopted by Cooper et al. (2010) to classify galaxies as red and blue (see Section \ref{subsec.env.uncertainty}). The fraction of galaxies classified as red is shown in all panels.}
\label{fig.gr_multiplot}
\end{figure*}

Figure \ref{fig.gr_multiplot} shows the galaxy colour $(g-r)$ histograms at the rest-frame in absolute magnitude ($M_{\rm r}$) and environment contrast ($\log_{10}(1+\delta)$) bins between 0.01$<$$z$$<$0.4. There exists a noticeable relation between the galaxy environment, luminosity and fraction of red galaxies. For a given luminosity bin, the fraction of red galaxies increases as the environment becomes denser. The blue cloud becomes less prominent and the red sequence more evident as the local density increases for all luminosity bins. Red galaxies are only dominant ($f_{\rm red}$$>$0.5) for denser and more luminous objects, i.e., in the specific bin containing galaxies within $\log_{10}(1+\delta)$$>$+0.05 and $M_{\rm r}$$<$-21. Other works in the literature also found fractions of red and quiescent galaxies, mostly massive ones, between 0.55 and 0.7 in typical galaxy cluster halos and galaxy pairs \citep[see][]{vanderWeletal2010, Pattonetal2011}.

% Spearman coefficient
%\begin{equation}
%r_s = \frac{\sum_i (R_i - \bar{R})(S_i - \bar{S})}{\sqrt{\sum_i (R_i - \bar{R})^2}{\sqrt{\sum_i (S_i - \bar{S})^2}}}
%\end{equation}

\section{Galaxy Population in G3C groups}
\label{sec.g3c_groups_results}

Since the G3C groups have been identified from a magnitude-limited sample, any galaxy population analysis would demand a strong selection function correction by using only GAMA data. The KiDS/DR3 volume-limited sample is then suitable to carry out a homogeneous analysis of the group sample, keeping the same selection function (or luminosity threshold of galaxies) over all galaxy systems. 

Our analysis considers GAMA galaxies in G3C groups  ($r$<19.8) and on their outskirts (up to twice the radius that contains 100\% of all group members, i.e., 2$R_{100}$) within the group velocity dispersion ($\sigma_{\rm group}$), previously calculated by \cite{Robothametal2011}. KiDS galaxies around groups are extracted from the KiDS volume-limited sample following the redshift uncertainty of the photo-zs, selecting galaxies in the redshift range $z_{\rm group}\pm 0.042(1+z_{\rm group})$ within $R$$\le$$2R_{100}$ around the structure center. In summary, this analysis consists of combining the shallower spectroscopic sample from GAMA and a deeper and volume-limited samples from the KiDS database in order to keep the homogeneity of galaxy population in all groups within the redshift range. Essentially, it considerably increases the redshift range of our analysis which would be much smaller if we only consider spectroscopic data.    

\subsection{The galaxy environment in G3C groups}

\begin{figure*}
\centering
\includegraphics[scale=0.27]{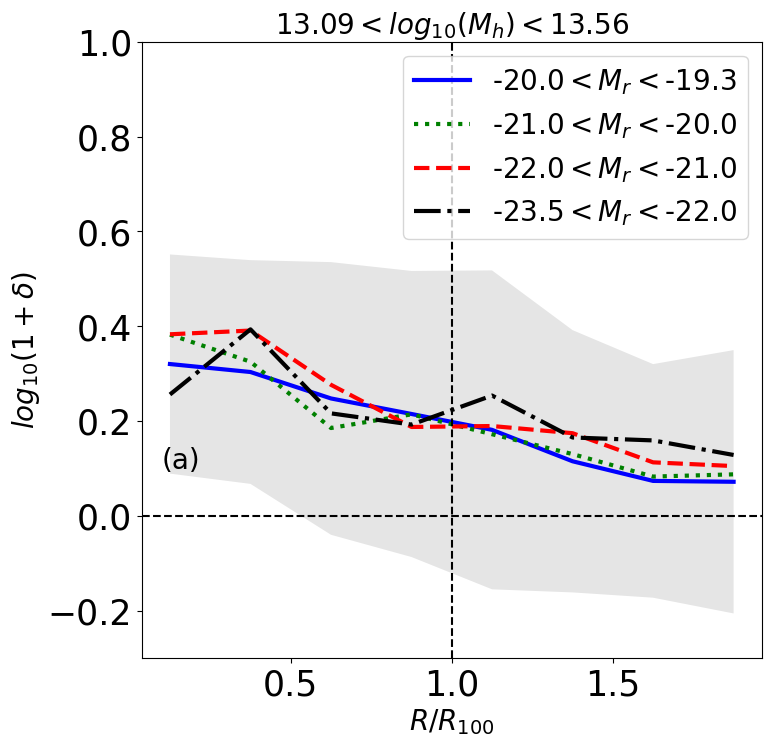}
\includegraphics[scale=0.27]{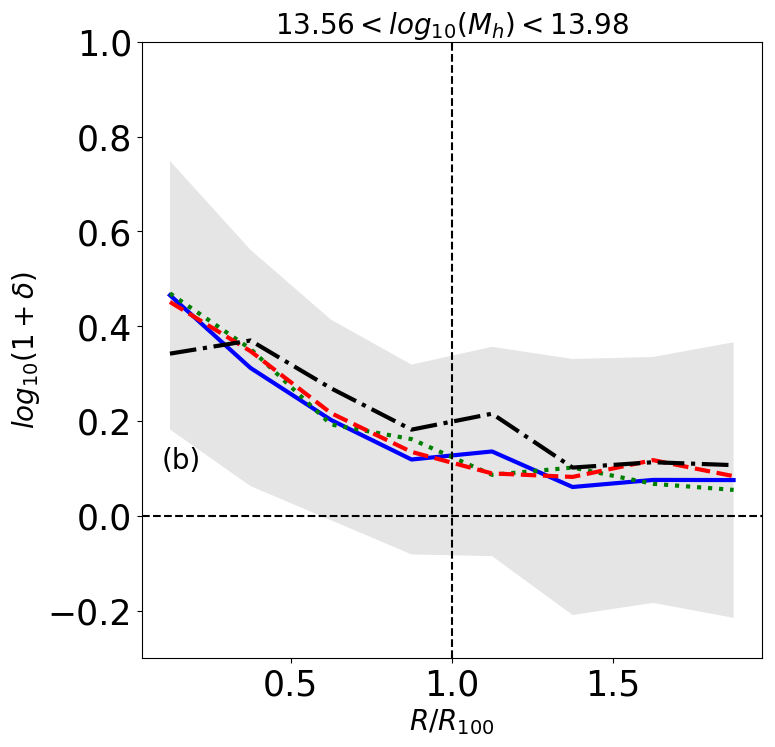}
\includegraphics[scale=0.27]{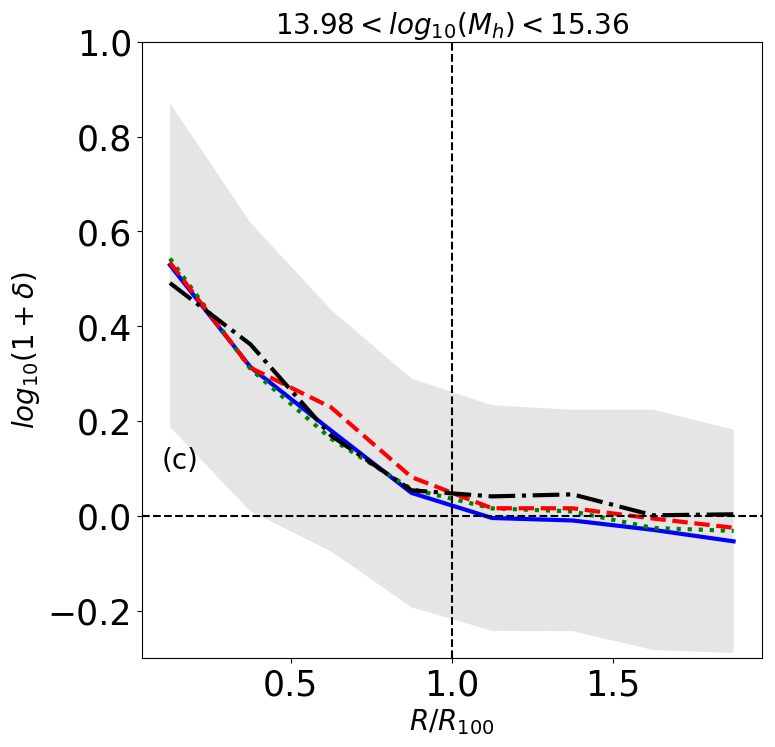}

\caption{The density contrast ($\log_{10}(1+\delta)$) as a function of the normalised radius of G3C groups ($R/R_{100}$). The luminosity bins M1, M2, M3 and M4 are represented by blue, green, red and black lines, respectively. The mean gradients of increasing G3C halo mass bins are represented by the panels from (a) to (c), respectively. The vertical and horizontal lines represent the normalised radius of the group (R/R$_{100}$=1) and the median density of the KiDS/DR3 galaxy sample, respectively. The shaded area represents 1$\sigma$ uncertainties and the dashed line represents the one unit of group radius.}
\label{fig.delta_log10rho_radius}
\end{figure*}

The galaxy environment is evaluated in G3C groups as a function of the normalised group radius (R/R$_{100}$) and absolute magnitude bins. Figure \ref{fig.delta_log10rho_radius} shows the density contrast as a function of the normalised radius compared to the central values of G3C groups for galaxy luminosity bins and different group mass ranges. The median gradient for the lowest mass group bin has values of +0.4 dex at central cores and +0.1 dex at outer regions (R/R$_{100}$$>$1). There is no statistical difference between the gradients with distinct luminosity bins according to the mean error bar (shaded area). This is probably related to the limitation of the data due to photo-z uncertainties. As the group mass increases, the amplitude of the gradients increases significantly. High mass groups gradients show values around +0.6 at central regions and +0.1 dex on the outskirts, showing the highest dense environment at the center of high mass galaxy systems. The density gradient becomes steeper and less luminosity-dependent for higher group mass bins. 

\subsection{The red/blue ratio of galaxies in G3C groups}

\begin{figure*}
\centering
\includegraphics[scale=0.3]{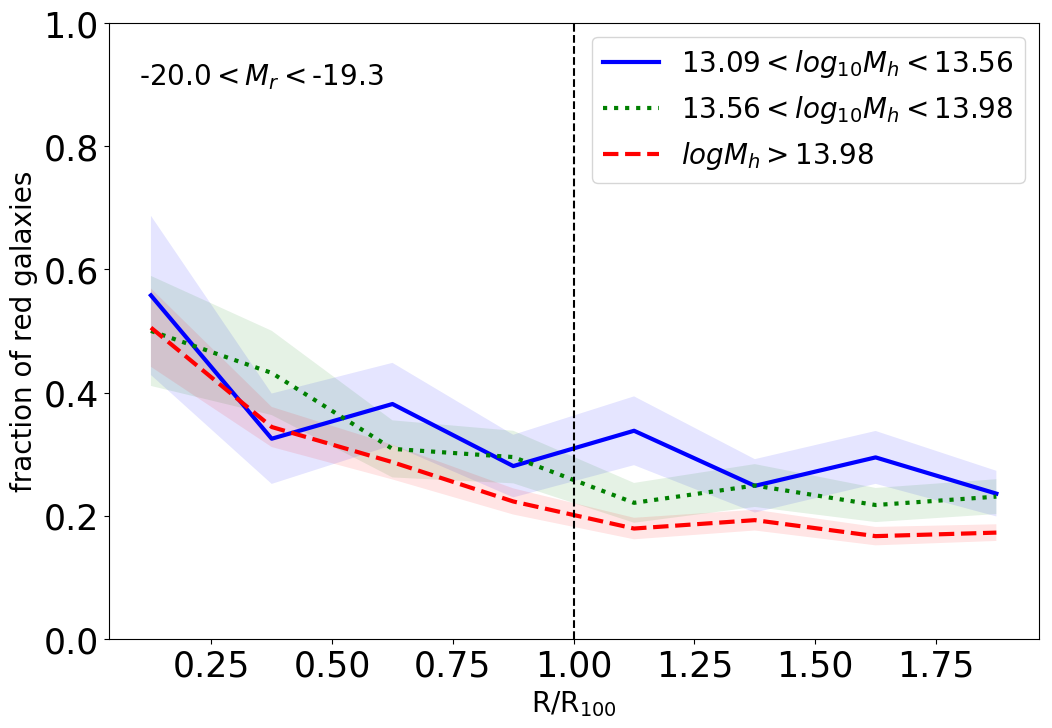}
\includegraphics[scale=0.3]{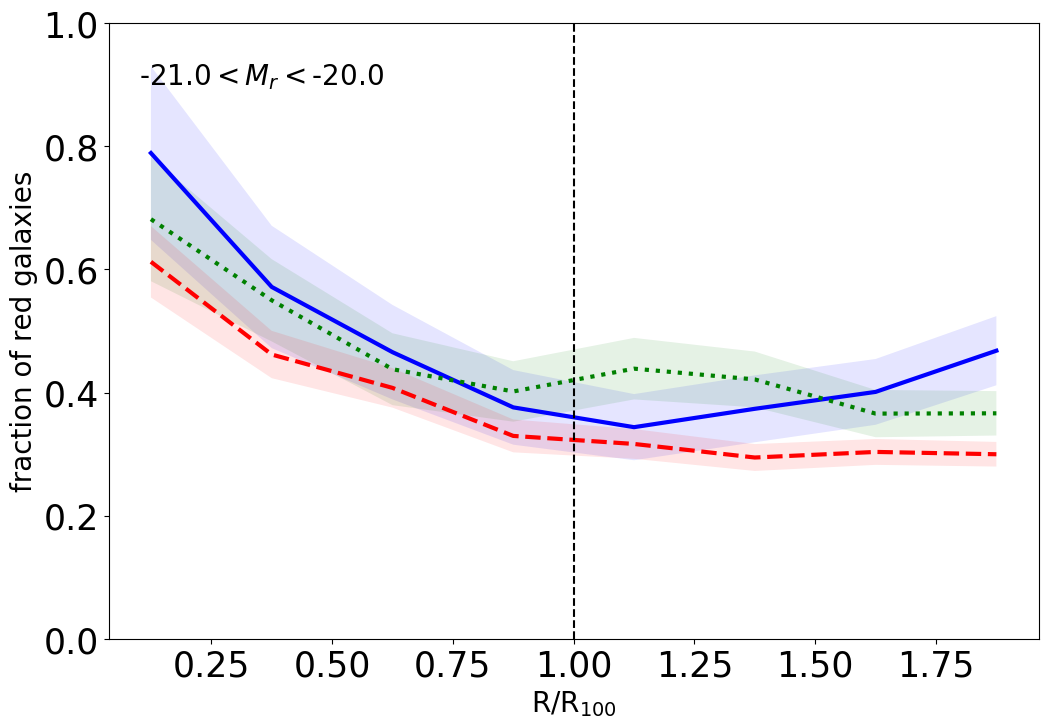}

\includegraphics[scale=0.3]{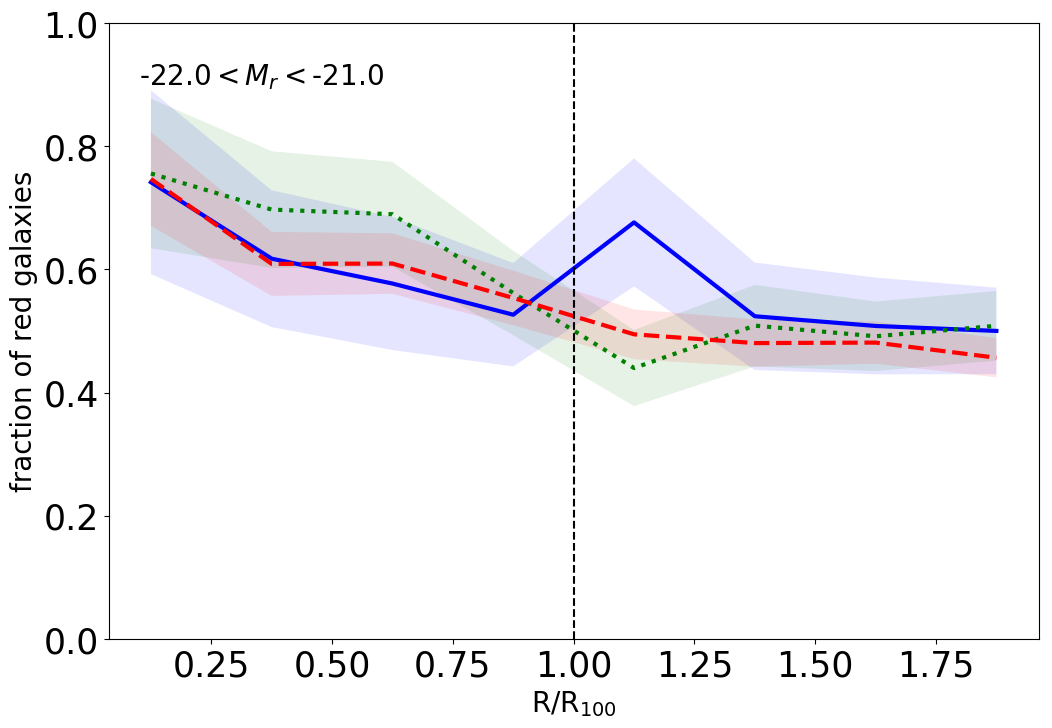}
\includegraphics[scale=0.3]{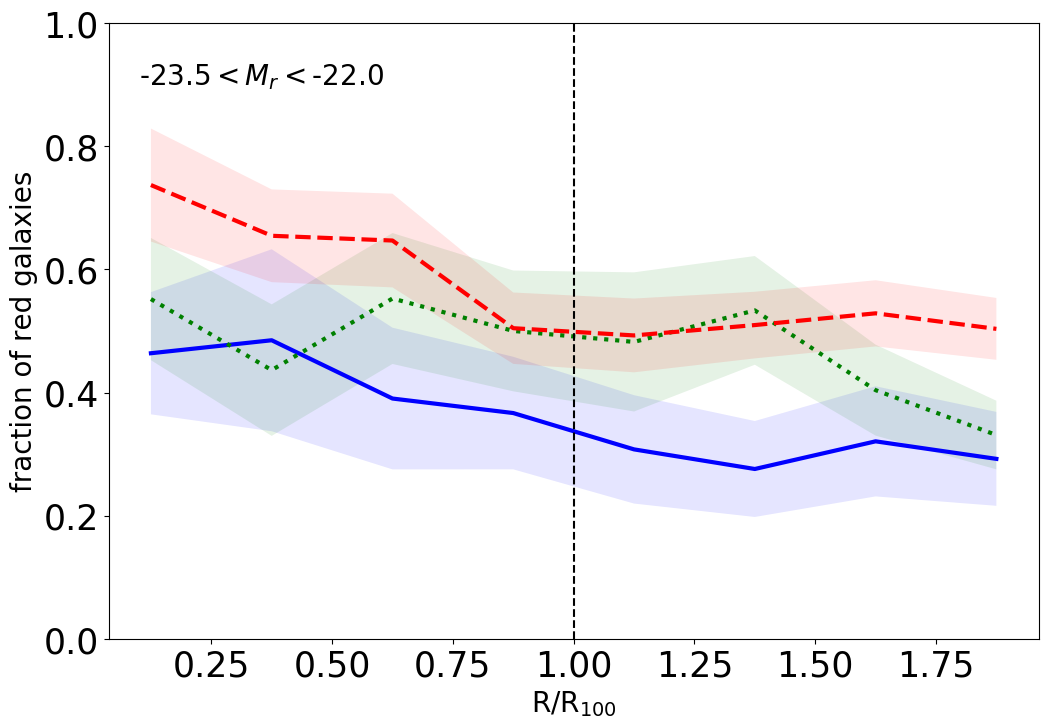}

\caption{The red galaxy fraction as a function of the normalised group radius (R/R$_{100}$) is shown for all luminosity bins. The red galaxy fraction gradients are shown for different G3C group mass bins. Shaded areas represent 1$\sigma$ dispersion.}
\label{fig.fred_Mabs_radius}
\end{figure*}

Figure \ref{fig.fred_Mabs_radius} shows the fraction of red galaxies as a function of the normalised radius of galaxy groups for galaxy luminosity and group mass bins. The red galaxy fraction clearly decreases as a function of the normalised radius for most cases, as expected. Due to photo-z uncertainties, there is no significant difference between these relations, unless for the brightest galaxy luminosity bin, between the lowest and highest group mass bins. The faintest galaxies ($M_{\rm r}$$>$-20) present red galaxy fractions around 0.5 at group cores and decrease on the outskirts, reaching values around 0.3. These radial gradients become redder and more prominent for more luminous bin. For the next two luminosity bins (-21<$<$$M_{\rm r}$$<$-20 and -22$<$$M_{\rm r}$$<$-21), the red galaxy fractions at the inner is 0.7 and at outer regions reaches 0.4. The brightest luminosity bin shows higher dispersion between group gradients, when compared to other luminosity bins. Only for the brightest luminosity bin, it is possible to differentiate the lowest and highest mass group gradients. Besides the high dispersion due to the low number of galaxies, the fraction of red galaxies is systematically higher for high mass groups than for low ones. Figure \ref{fig.fred_Mabs_radius} points out the group environment specifically acts on low and intermediate luminosity galaxies, indicated by the gradient of red galaxies as a function of the radius. This gradient becomes less remarkable as the galaxy luminosity increases. For the most luminous galaxies ($M_{\rm r}$$<$-22), its slope gradient is smaller when compared to other bins but it is still statistically detected. Studies based on the SDSS galaxy groups showed that central and luminous galaxies presents slight correlation between the red fraction and environment \citep{Tinkeretal2011}. On the other hand, the correlation between ($g$-$r$) versus environment is more pronounced among satellite galaxies \citep{Tinkeretal2017, Dracomiretal2017}. These results are confirmed here by showing a steeper gradient for objects between $-22.<M_{\rm r}<-20.$ when compared to the brightest galaxies. 

\begin{figure*}
\centering
\includegraphics[scale=0.22]{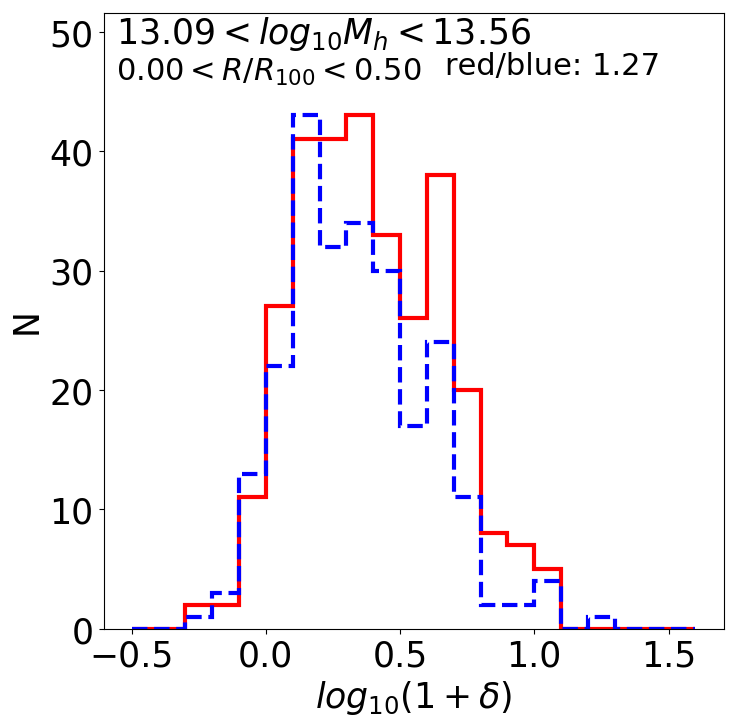}
\includegraphics[scale=0.22]{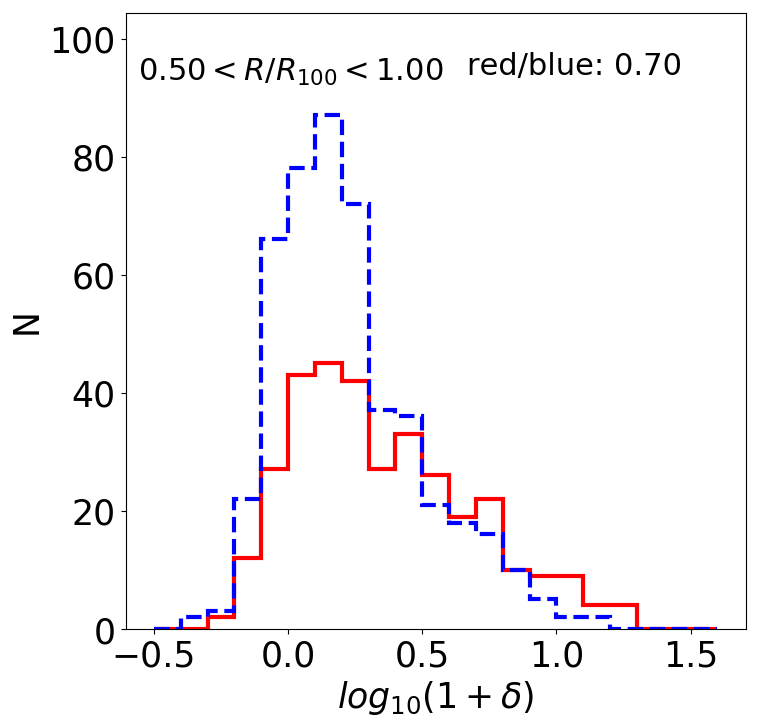}
\includegraphics[scale=0.22]{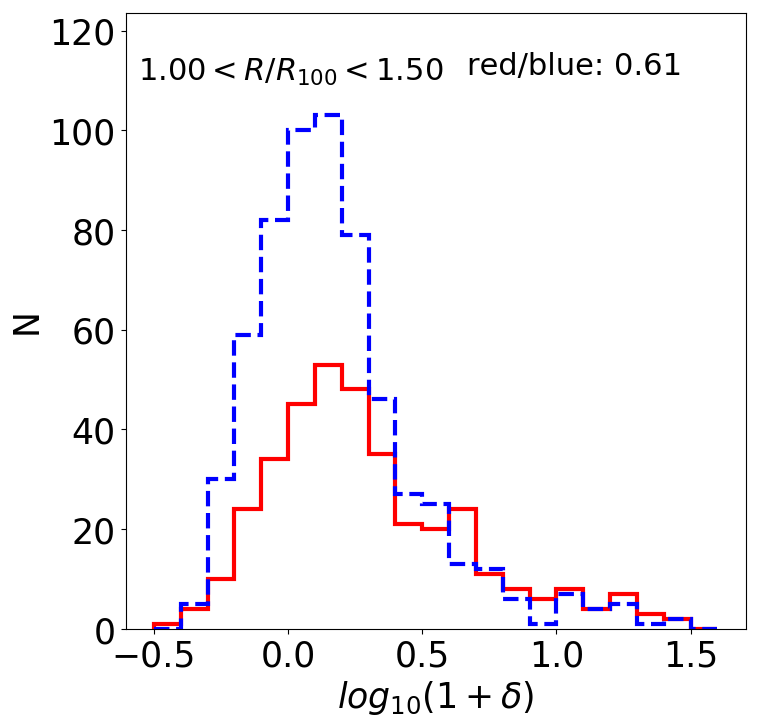}
\includegraphics[scale=0.22]{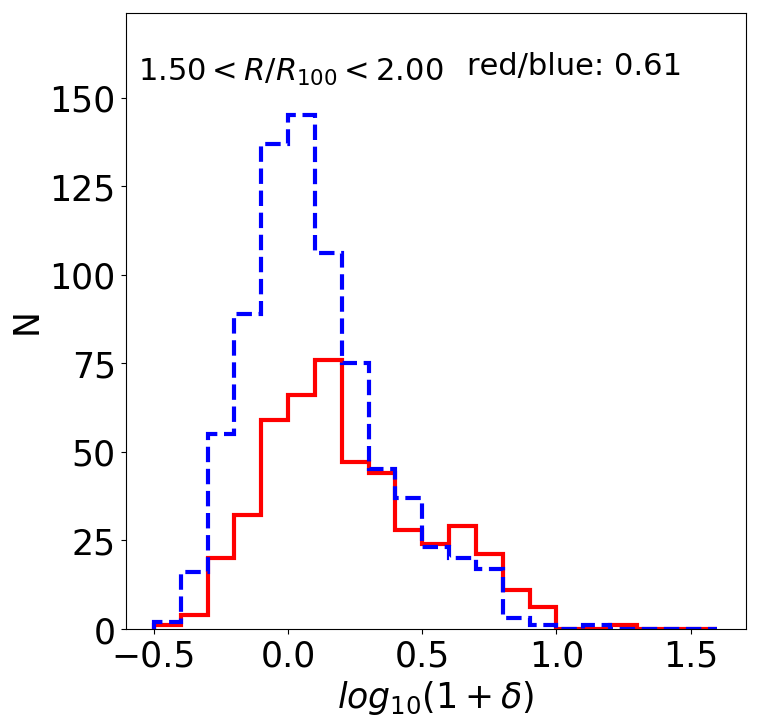}

\includegraphics[scale=0.22]{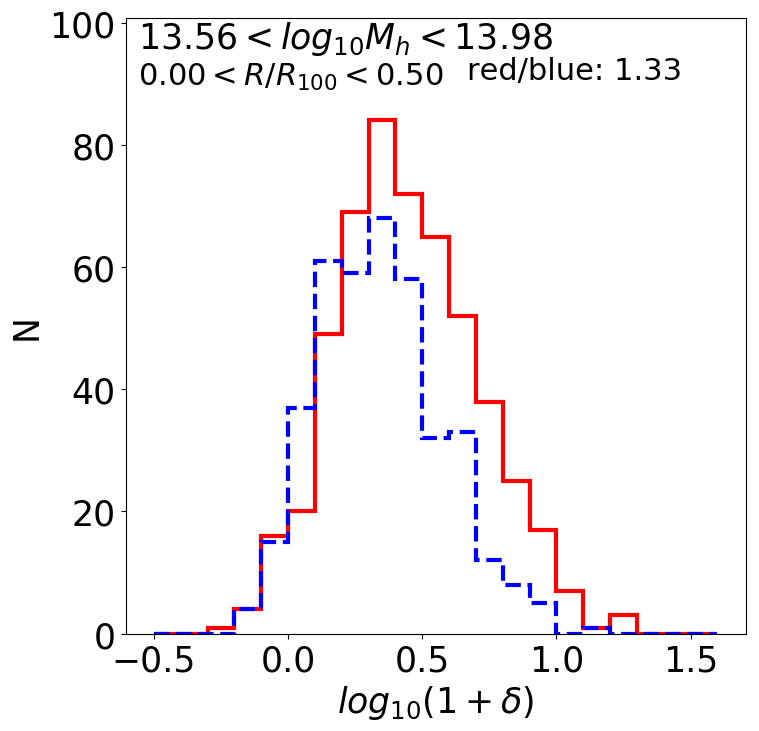}
\includegraphics[scale=0.22]{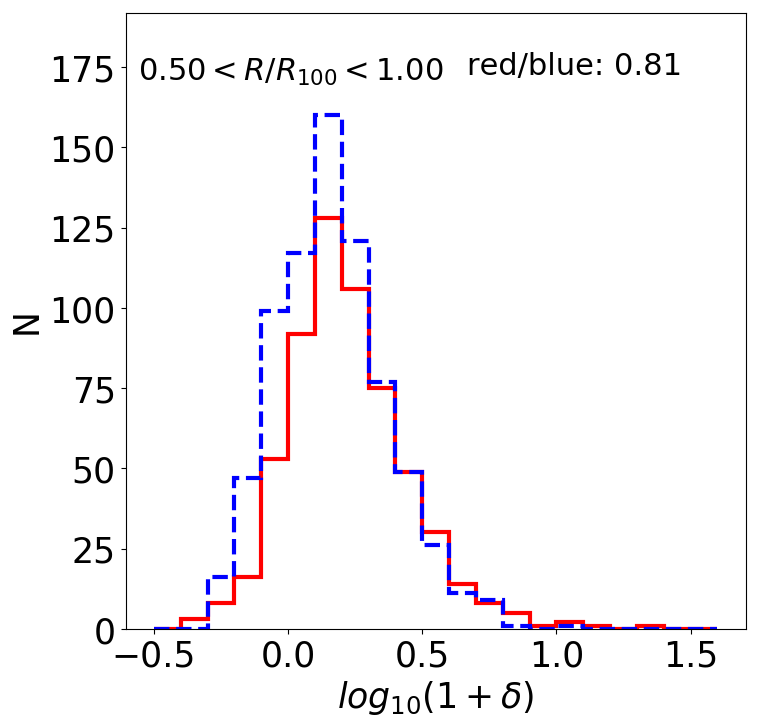}
\includegraphics[scale=0.22]{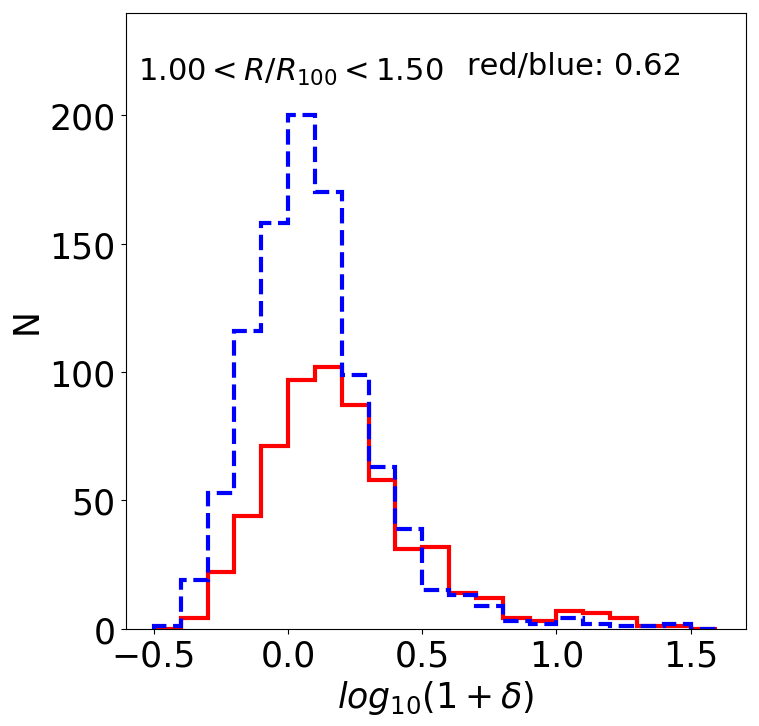}
\includegraphics[scale=0.22]{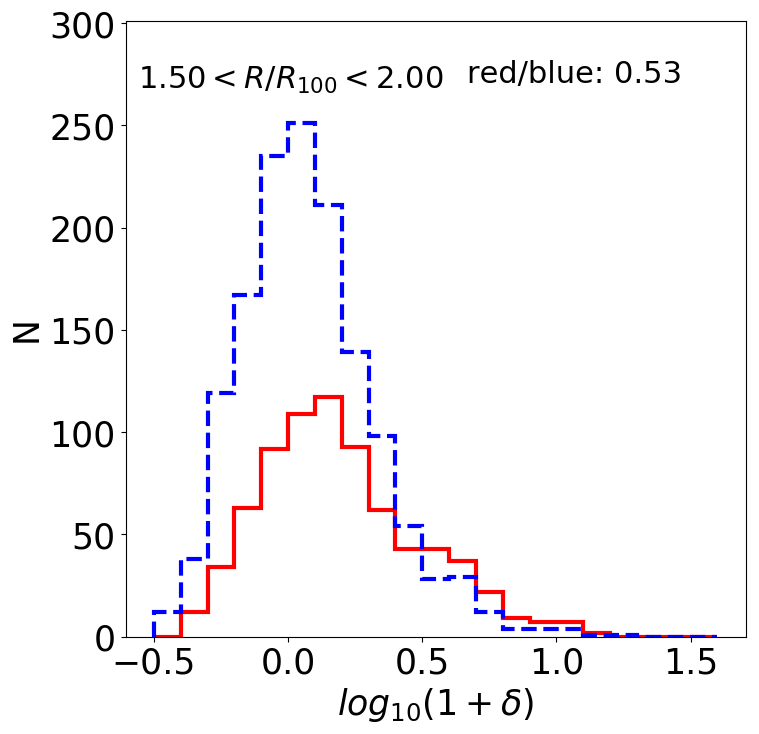}

\includegraphics[scale=0.22]{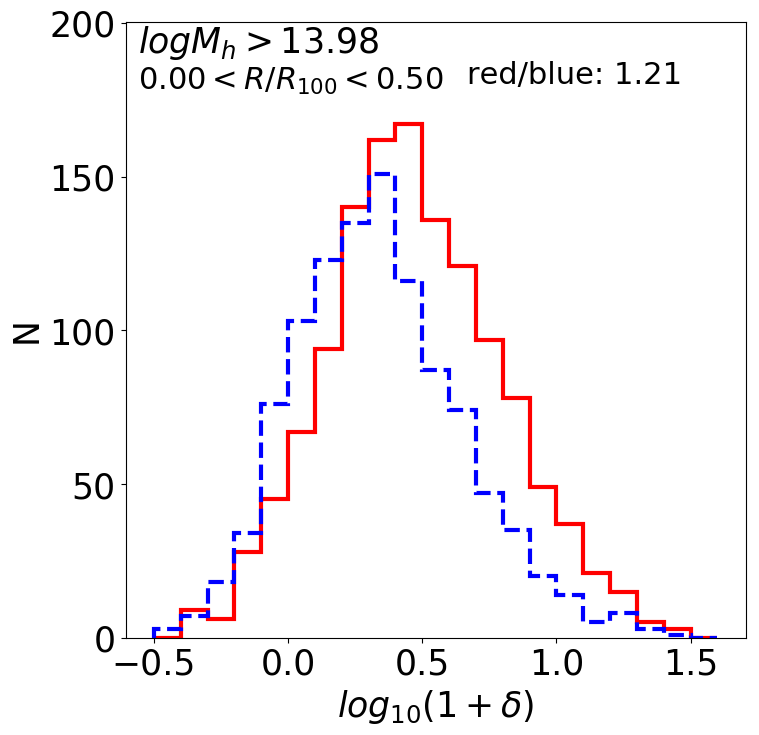}
\includegraphics[scale=0.22]{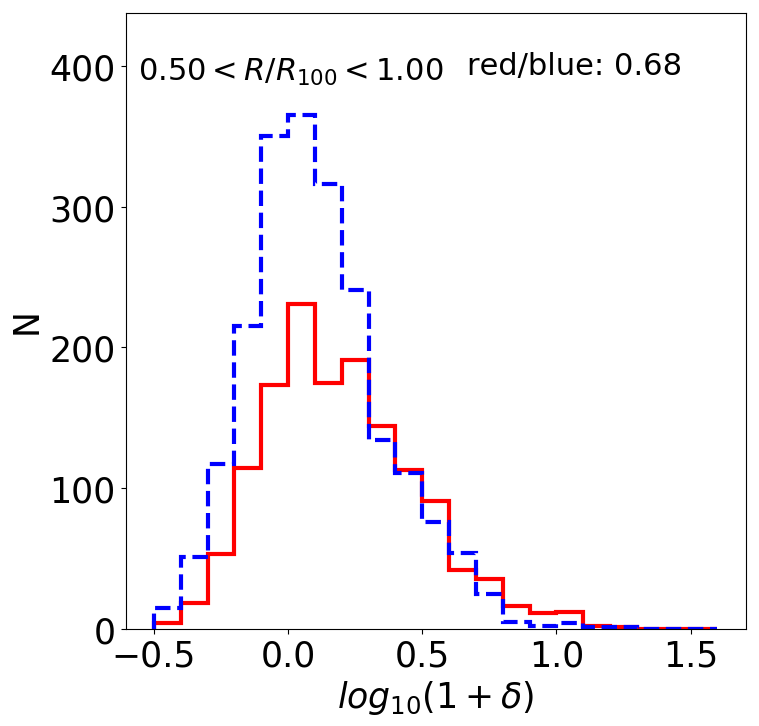}
\includegraphics[scale=0.22]{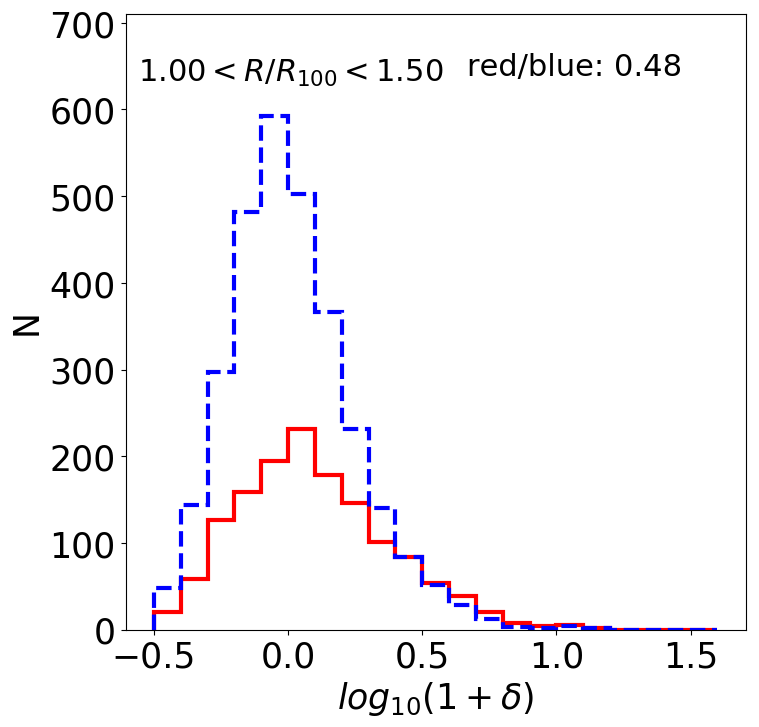}
\includegraphics[scale=0.22]{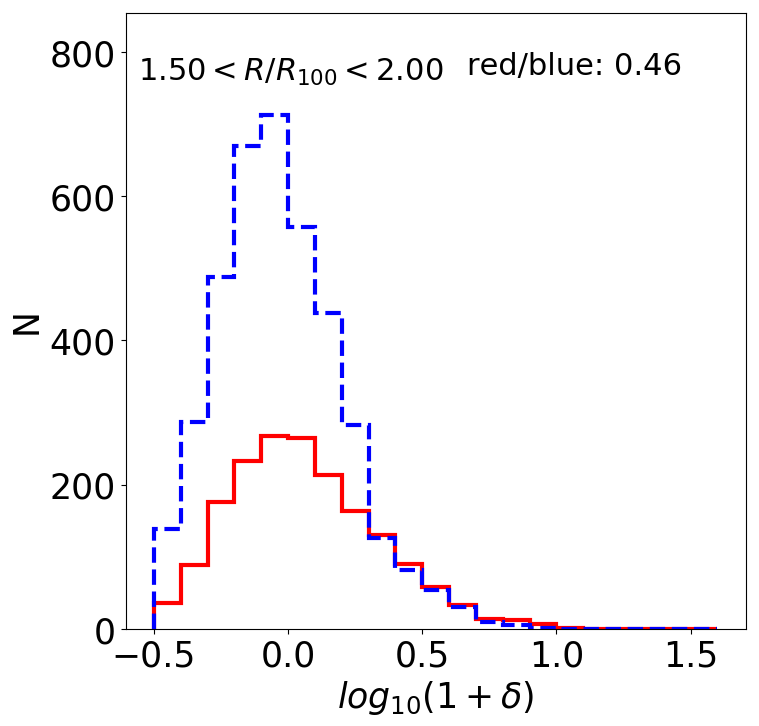}

\caption{The density contrast distributions of galaxies classified as red (solid line) and blue (dashed line) shown in group radius bins and group mass bins. The radius bins are defined as 0.5 $R/R_{100}$ wide, up twice the normalised radius. The solid and dashed lines represent the red and blue galaxy distributions, respectively. The fraction of red and blue galaxies are shown at the upper region of all panels.}
\label{fig.hist_log_delta_red_blue}
\end{figure*}

Figure \ref{fig.hist_log_delta_red_blue} shows the density contrast distribution of galaxies classified as red and blue as a function of the distance from the group centers, separated by group mass bins. We define four bins in group radius between group centers and 2$R/R_{100}$. Comparing the distributions of the density contrasts of blue and red galaxies at the central regions of groups, we notice a high density regions excess for red galaxy distributions when compare with the blue ones, particularly for high mass groups. The dominance of red galaxies over blue ones is also evident in all group mass bins between 0$<$$R/R_{100}$$<$0.5, reaching a fraction red/blue $\ge1.21$. As the radius increases, the high density excesses noticed for red galaxy distributions are not evident anymore. Furthermore, the fraction of red/blue objects decreases as the normalised radius increases and reaches the ratios around 0.5. There is also an excess of high density regions on the outskirts of lower mass groups, particularly at 1.0$<\log_{10}(1+\delta)$$<$1.5. In low mass groups, the task of center definition is more susceptible to miscentering because of the low number of galaxy members. This effect is noticeable for both galaxy populations so it might be a geometrical effect instead of a dynamical state of the galaxy systems. This effect does not depend on galaxy population but only on group mass.  

\subsection{The contamination in group gradients due to redshift uncertainties}
\label{contamination_redshift}

One of the consequences of combining spec-z and photo-z in our analysis is the projection effects. The gradient analysis from the galaxy groups shown here is based on the projected sky plane. Group members close to the group centers in the 2D sky plane can actually be background or foreground objects in the redshift range $z_{\rm group}\pm 0.042(1+z_{\rm group})$. Due to the photo-z uncertainties, it is not possible to deproject these objects. Our gradients are calculated using galaxies in the sky plane and within $z_{\rm group}\pm0.042(1+z_{\rm group})$. This projection effect systematically decreases the discrepancies between the galaxy populations and the density contrast gradients due to background and foreground contaminations. On the other hand, this projection effect is homogeneous over all G3C groups since our gradients follow the redshift uncertainties of KiDS galaxies over all redshift bins, keeping the same background contamination throughout the redshift range. 

Having this limitation in mind, we also analysed galaxy haloes above $\log_{10}(M_{FoF}/M_{\odot})=13$ in our mock catalogue (see Section \ref{subsec.mock_cat}) by using the same criteria of group gradients aforementioned. The results indicated that the fraction of halo members recovered and the contamination of foreground and background galaxies due to photo-z uncertainties is 67.3\% and 48.4\% for the KiDS photo-z uncertainties, respectively.

In addition, comparing the fraction of red galaxies with other works in the literature, our red galaxy fraction is similar to those found by \cite{vanderWeletal2010} (see their Figure 1) at central regions of rich clusters ($\log_{10}(M_{\rm h})\sim15$), between 0.6 and 0.8. It is important to mention that we are aware of projected galaxies in the line-of-sight due to photo-z uncertainties and its effects, however, our results are always shown in a comparative way, separating the galaxy sample into luminosity, distance from the group center and group masses.

\subsection{The influence of group center definition}
\label{sec.center}

We initially adopt the BCGs as centers of G3C groups (Section \ref{subsec.gama_g3c}), as previously shown. Nonetheless, the GAMA G3C catalog also provides other group center definitions for the galaxy systems. Thus, we evaluate the influence of a second center definition in our results. The $r$-band luminosity-weighted center is then employed to evaluate how sensitive the density contrast distributions of red and blue galaxies are to the center definition. The Appendix \ref{app.sec_center_influence} illustrates the same analysis as shown in Figure \ref{fig.hist_log_delta_red_blue} but now using the $r$-band luminosity-weighted group centers. We notice that the red/blue fractions and the red and blue histograms are similar to the ones calculated using the BCG as definition of group center. Consequently, our conclusions are insensitive to the new center definition. The local density excess found on the outskirts of low mass systems is still found at $\log_{10}(1+\delta)$$>$1.0 (see Section \ref{contamination_redshift}). The standard deviation of the offsets between both group center definitions ($|\mathbf{r}_L$ - $\mathbf{r}_{BCG}|$) is $\sim$0.09h$^{-1}$Mpc, corresponding to 12\% of the average $R_{100}$ for our group sample. This relatively small offset indicates that the group center definitions for G3C groups are quite stable and do not change our previous conclusions. 

\section{Conclusions and Discussion}

We investigated the galaxy environment in GAMA G3C groups using a volume-limited galaxy sample ($M_{\rm r}$$<$-19.3 and 0.01$<z<$0.4) from the Kilo Degree Survey Data Release 3. The k-NN technique (k-Nearest Neighbour) was adapted to take into account the adapted photo-z PDFs of galaxies in the galaxy environment calculations. In order to evaluate the performance of our adapted technique, we generated a simulated volume-limited sample ordered in tiles. These tiles also mimic the sky regions affected by masking and bad pixels. For the galaxy environment analysis in GAMA G3C groups, we selected G3C groups within the same redshift as the KiDS/DR3 galaxy sample, selecting halo masses $\log(M_h)>13.21$.

Our main findings are the following:

\begin{itemize}

\item The simulated KiDS/DR3 tiles showed the capability of the adapted k-NN technique to recover the galaxy environment in the KiDS/DR3 database. We were able to recover the relation between the galaxy environment, luminosity and the galaxy colour $(g-r)$ up to $z$=0.4.  

\item Using the KiDS galaxy sample, we evaluated the galaxy population in these galaxy systems and on their outskirts. Density contrast gradients were systematically steeper for more massive systems, reaching on average +0.6 dex higher than their outskirts. 

\item We separated the galaxy population into two main classes, blue and red ones using a colour-magnitude cut adopted by \cite{Cooperetal2010}. The fraction of red galaxies as a function of the normalised radius ($R/R_{100}$) presents, for the faintest galaxies, $\sim$50\% of red galaxies and decreases as the radius increases. As the luminosity increases, it reaches $\sim$80\% at group centers and decreases on the outskirts. Higher dispersion is noticed for the most luminous bin, probably due to the low number of galaxies.  

\item The density contrast distribution for red galaxies showed an excess of high density regions when compared to the blue galaxies at the center of groups ($R/R_{100}$$<$0.5). The dominance of red galaxies was also noticed at the central part of these systems. In contrast to the red one, the blue distribution was dominant at the outer regions of the groups and beyond their central cores ($R/R_{100}$$>$0.5). The red/blue fraction decreases as the normalised radius increases, reaching values around 50\%. 

\item The influence of the group center definition on our results is also evaluated. First, the brightest cluster galaxy as center definition is employed for our main conclusions. Using the r-band luminosity weighted center as a new center definition, similar conclusions pointed out the insensitivity of the center definition in our analysis.    

\end{itemize}

Several mechanisms can be responsible for the galaxy quenching found in this work, acting on galaxies at different distances from the group center, such as merging \citep{Icke1985, Mihos1995} and harassment \citep{Mooreetal1996, Mooreetal1999} over all scales, and ram-pressure \citep{GunnGott1972} and tidal-stripping \citep{Nulsen1982, ToniazzoSchindler2001} at the inner regions. The correlation between the fraction of red galaxies and the local density was previously found in the literature, being lower fractions for fainter galaxies \citep[e.g.][]{Balletal2008}. However, this result is not found here probably due to the photo-z uncertainties of the KiDS database. The current quenching scenario predicts that hydrodynamical quenching mechanisms (e.g. ram-pressure) slowly remove the cold gas from galaxy halos and consequently quench the infall galaxy. An abrupt and extreme quenching mechanism (mechanical ones, such as mergers or harassment) would perturb the gas within the galaxy halo and then trigger the star formation in these galaxies. As a consequence, it would reduce the fraction of the red galaxies. Hydrodynamical effects are mainly responsible for the smoothly colour changes at the outer part of galaxy groups and clusters. The intra-cluster hot gas is the main candidate to carry out this hydrodynamical quenching at that region. Recently, \cite{Zingeretal2016} used simulations to propose that the quenching process starts much earlier, beyond the virial radius and its consequences are only observed 2-3 Gyrs after the initial quenching. Another explanation can be the "splashback" galaxies. Having a highly excentric orbit, spiral galaxies in infall process would rapidly pass through the inner virial radius of the cluster and lose their neutral hydrogen. After that, they are already in quenching process and will spend most of the time on the cluster outskirts (1-2.5 virial radii) due to their eccentric orbits \citep{Mamonetal2004}.

At the inner parts, the mechanical processes are responsible for perturbing galaxies, often causing morphological transformation \citep[e.g.][]{vonderLindenetal2010}. In summary, there is no specific mechanism that fully explains both colour-environment and morphology-environment relations in galaxy clusters. They act all together in order to reproduce the observed transition from disky/star-forming galaxies to spheroidal/passive ones \citep{ParkHwang2009}.

The galaxy environment technique presented here can be also applied on other galaxy surveys in the future, such as S-PLUS (Mendes de Oliveira et al., in preparation), J-PLUS (Cenarro et al., in preparation), J-PAS \citep{Benitezetal2014} and EUCLID \citep{Clemensetal2015}.   

\section*{Acknowledgements}

MVCD thanks the financial support from FAPESP (processes 2014/18632-6 and 2016/05254-9) and the University of Leiden, the Netherlands, for their hospitality.
AM acknowledges the financial support of the Brazilian funding agency FAPESP (Post-doc fellowship - process number 2014/11806-9) 
MB is supported by the Netherlands Organization for Scientific Research, NWO, through grant number 614.001.451.
This work has made use of the computing facilities of the Laboratory of Astroinformatics (IAG/USP, NAT/Unicsul), whose purchase was made possible by the Brazilian agency FAPESP (grant 2009/54006-4) and the INCT-A.
GVK acknowledges financial support from the Netherlands Research School for Astronomy (NOVA) and Target. Target is supported by Samenwerkingsverband Noord Nederland, European fund for regional development, Dutch Ministry of economic affairs, Pieken in de Delta, Provinces of Groningen and Drenthe.

\appendix

\section{The influence of the group center defintion}
\label{app.sec_center_influence}

\begin{figure*}
\centering
\includegraphics[scale=0.22]{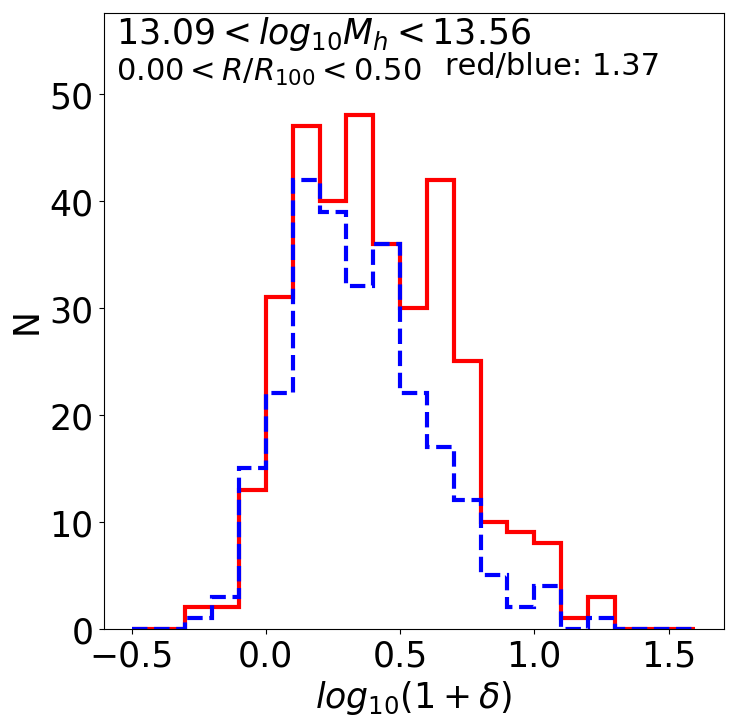}
\includegraphics[scale=0.22]{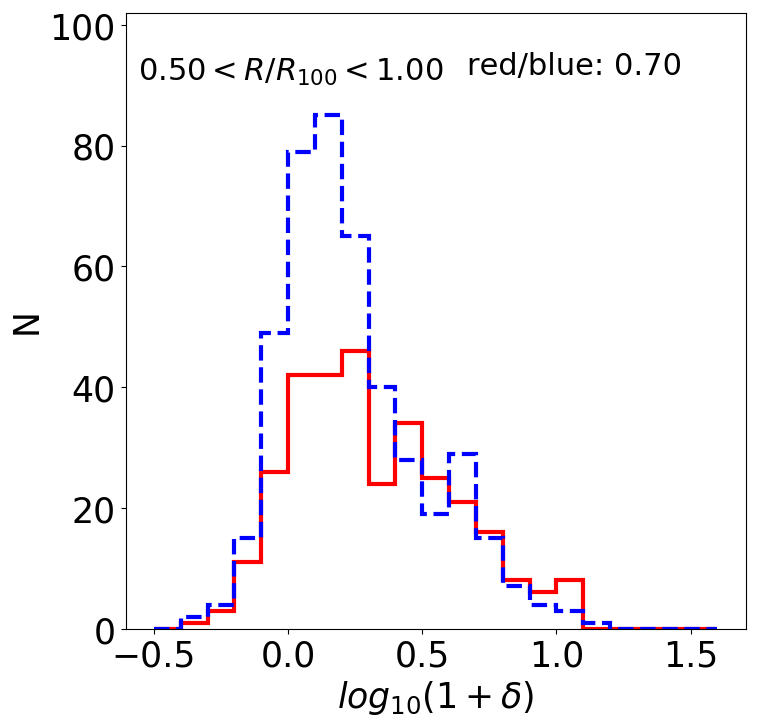}
\includegraphics[scale=0.22]{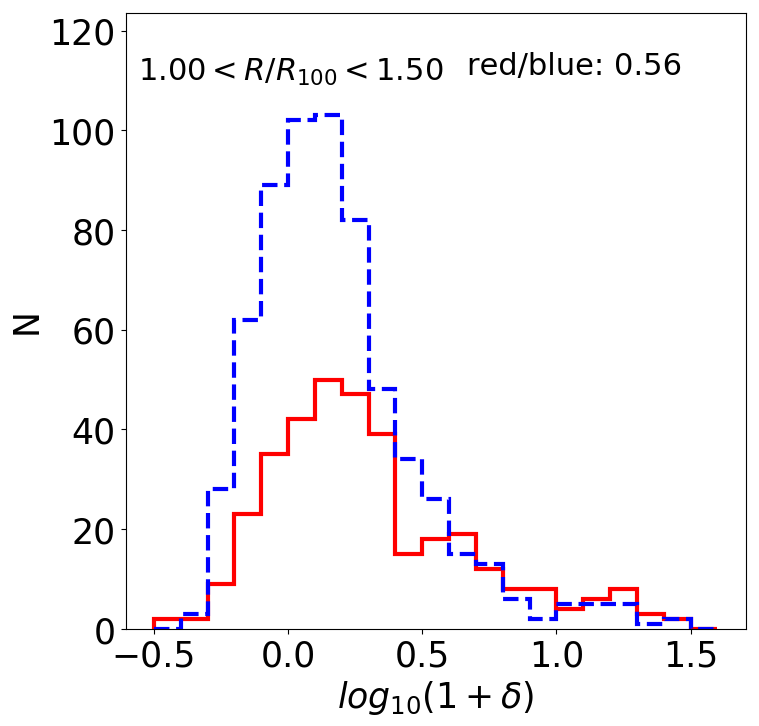}
\includegraphics[scale=0.22]{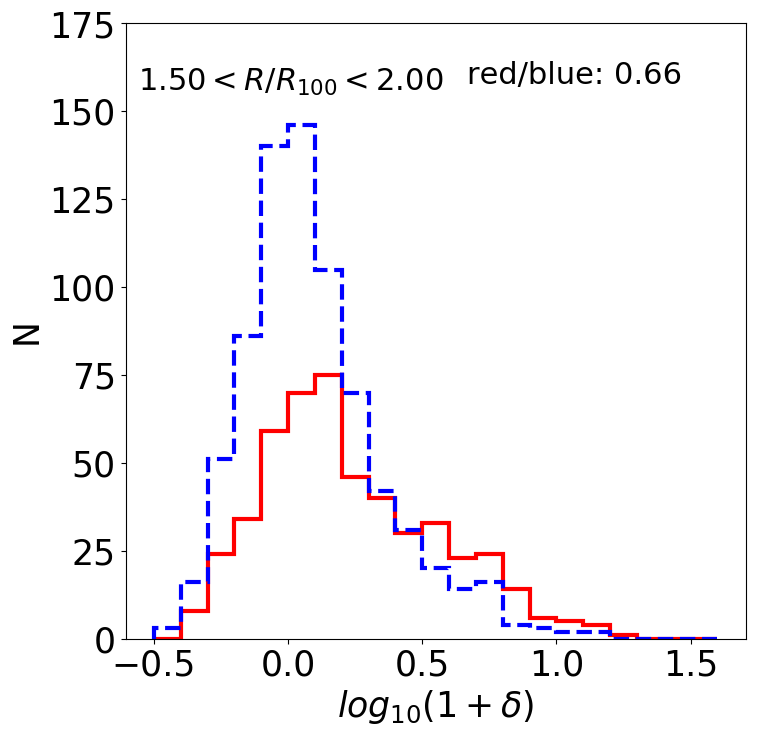}

\includegraphics[scale=0.22]{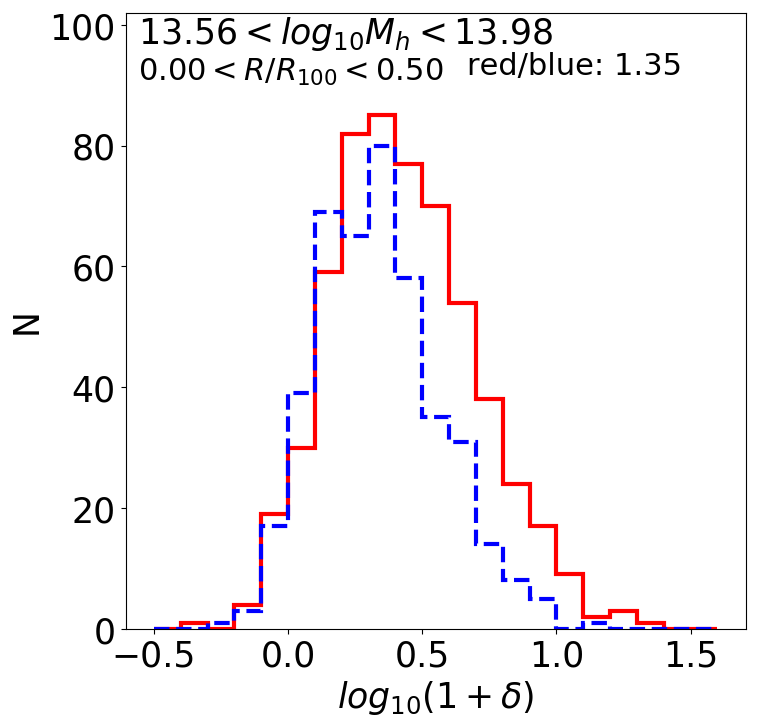}
\includegraphics[scale=0.22]{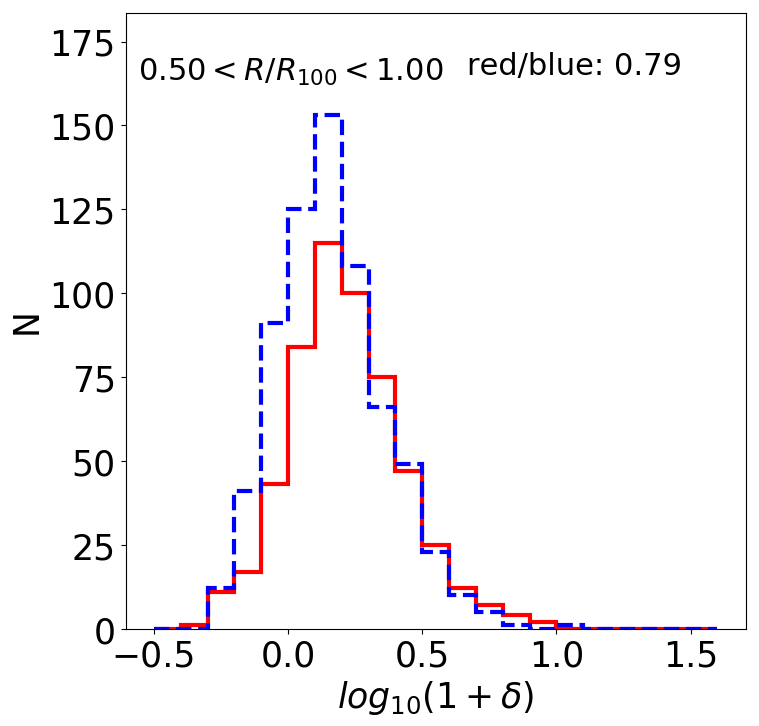}
\includegraphics[scale=0.22]{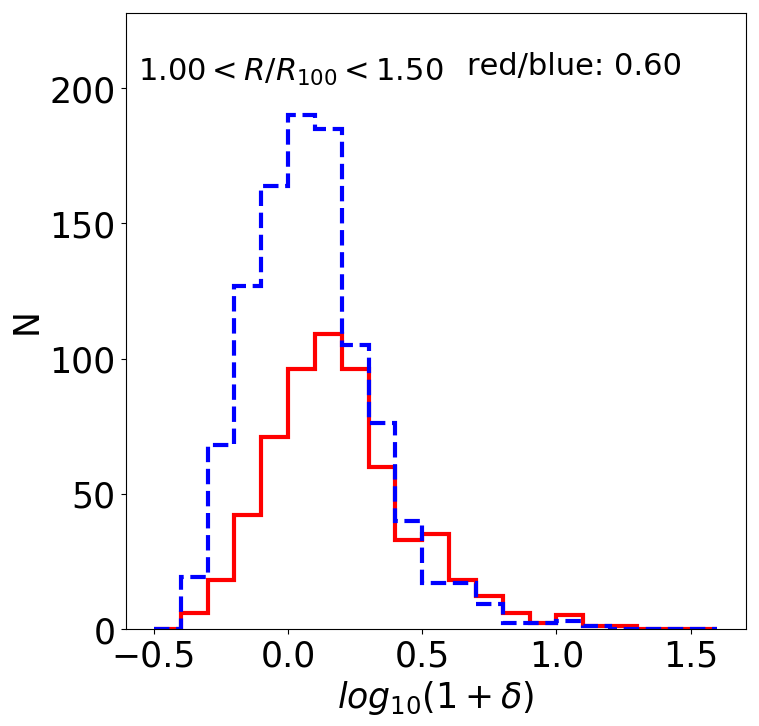}
\includegraphics[scale=0.22]{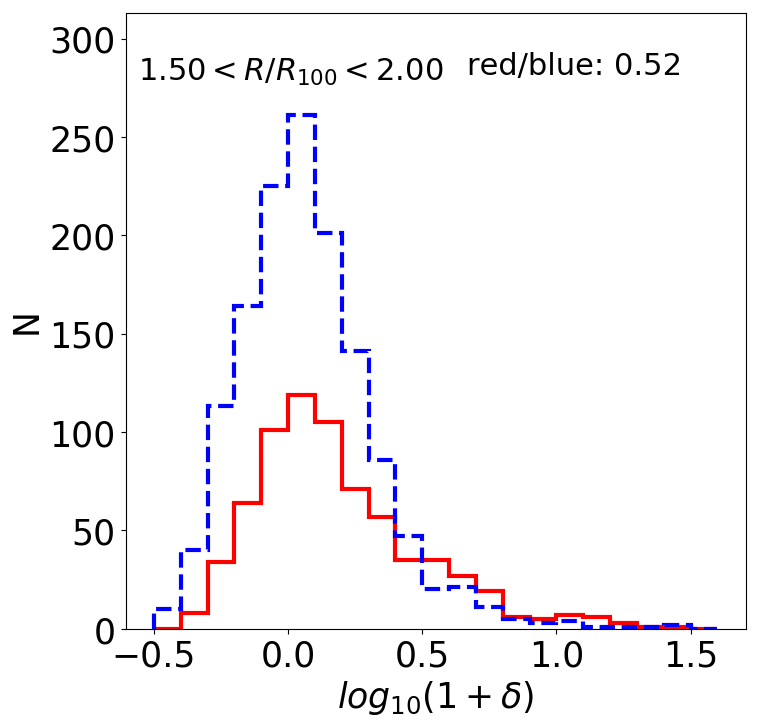}

\includegraphics[scale=0.22]{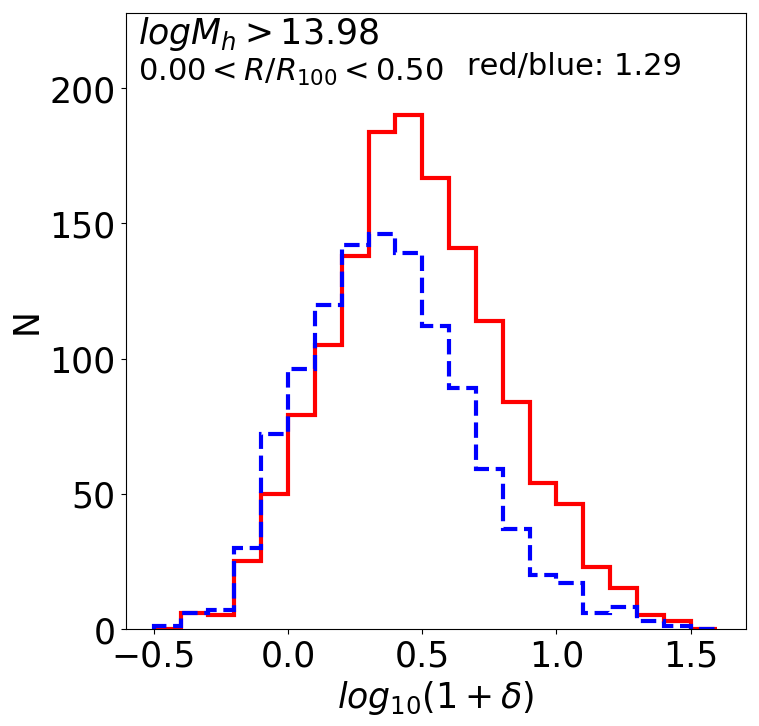}
\includegraphics[scale=0.22]{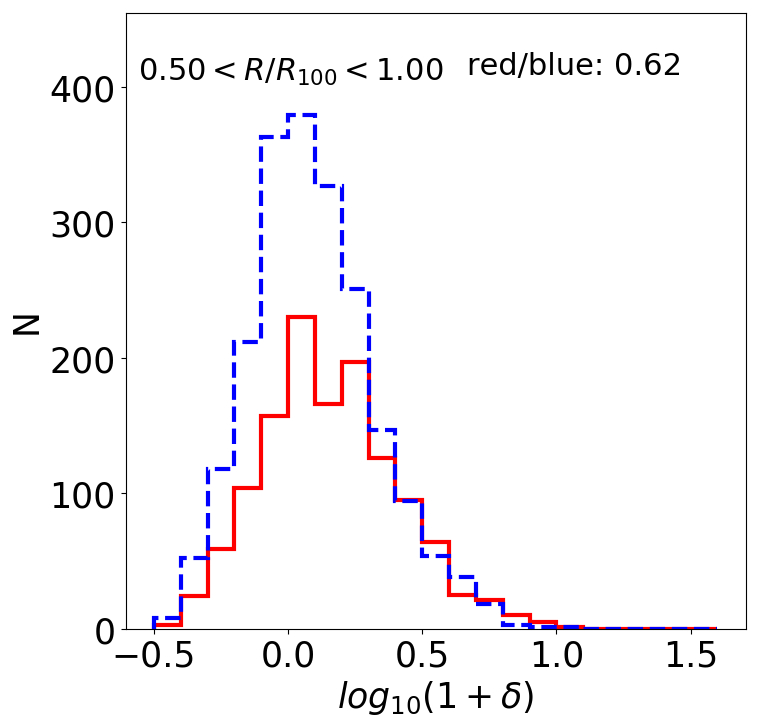}
\includegraphics[scale=0.22]{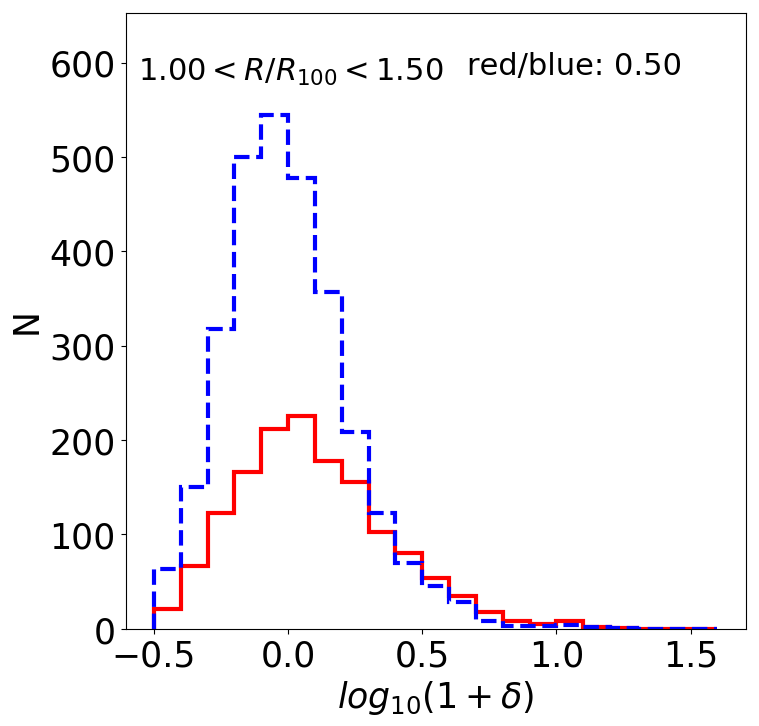}
\includegraphics[scale=0.22]{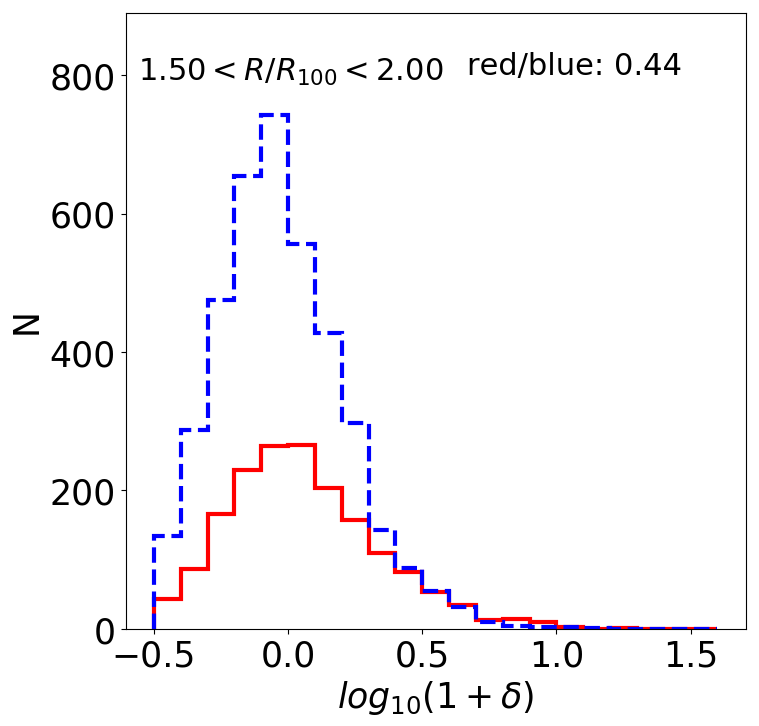}

\caption{The same as Figure \ref{fig.hist_log_delta_red_blue} but using the $r$ band luminosity weighted center defined by Robotham et al. (2011).}
\label{fig.hist_log_delta_red_blue_cen}
\end{figure*}

Figure \ref{fig.hist_log_delta_red_blue_cen} shows the density contrast distributions of galaxies classified as blue and red as a function of group mass and normalised radius bins for the r-band luminosity weighted group center. The comparison between Figures \ref{fig.hist_log_delta_red_blue_cen} and \ref{fig.hist_log_delta_red_blue} indicates that the center definition does not change our conclusions. Moreover, the conclusions obtained from Figures \ref{fig.delta_log10rho_radius} and \ref{fig.fred_Mabs_radius} are not changed either. 

% Don't change these lines
\bsp	% typesetting comment
\label{lastpage}
\end{document}